\documentclass[letterpaper,10pt,twoside]{article}

\usepackage[utf8]{inputenc}  
\usepackage[english]{babel} 
\usepackage[english=american]{csquotes}
\MakeOuterQuote{"}
\usepackage[title,titletoc]{appendix}
\usepackage[nottoc]{tocbibind}

\usepackage{authblk}
\usepackage{lmodern}
\usepackage{amssymb,amsthm,mathtools} 
\usepackage{hyperref} 
\usepackage{geometry} 

\usepackage{graphicx}  
\graphicspath{{images/}}

\usepackage{tikz}
\usetikzlibrary{calc,decorations.markings}
\usetikzlibrary{decorations.pathmorphing}
\usetikzlibrary{decorations.markings}
\usetikzlibrary{arrows.meta,bending}

\newcommand{\im}{\mathrm{i}}

\newcommand{\R}{\mathbb{R}}
\newcommand{\C}{\mathbb{C}}
\newcommand{\defeq}{\coloneqq}
\newcommand{\tens}{\otimes}
\DeclareMathOperator{\id}{id}
\newcommand{\xd}{\mathrm{d}}
\newcommand{\xD}{\mathrm{D}}
\newcommand{\cH}{\mathcal{H}}
\newcommand{\ls}{\ell}
\newcommand{\ms}{\mathsf{m}}
\newcommand{\oR}{\overline{R}}

\newcommand{\Lp}{\mathrm{p}}
\newcommand{\Le}{\mathrm{e}}
\newcommand{\Lin}{\mathrm{in}}
\newcommand{\Lout}{\mathrm{out}}
\newcommand{\Lcin}{{\overline{\mathrm{in}}}}
\newcommand{\Lcout}{{\overline{\mathrm{out}}}}
\newcommand{\Lint}{\mathrm{int}}
\newcommand{\Lext}{\mathrm{ext}}
\newcommand{\Lini}{\mathrm{ini}}
\newcommand{\Lfin}{\mathrm{fin}}
\newcommand{\Lvac}{\mathrm{vac}}

\newcommand{\no}[1]{{:}#1{:}}
\newcommand{\rop}{\sigma^+}
\newcommand{\lop}{\sigma^-}

\newcommand{\sangle}{\mathsf{\Omega}}

\allowdisplaybreaks

\renewcommand{\thesubsection}{\Alph{subsection}.}

\begin{document}


\begin{titlepage}
\title{\textbf{Interaction of evanescent particles with an Unruh-DeWitt detector}}
\author[1]{Daniele Colosi\footnote{email: dcolosi@enesmorelia.unam.mx}}
\author[2]{Robert Oeckl\footnote{email: robert@matmor.unam.mx}}
\author[2,3]{Adamantia Zampeli\footnote{email: azampeli@matmor.unam.mx}}
\affil[1]{Escuela Nacional de Estudios Superiores, Unidad Morelia,
Universidad Nacional Autónoma de México,
C.P.~58190, Morelia, Michoacán, Mexico}
\affil[2]{Centro de Ciencias Matemáticas,
Universidad Nacional Autónoma de México,
C.P.~58190, Morelia, Michoacán, Mexico}
\affil[3]{International Center for Theory of Quantum Technologies, University of Gdańsk, Wita Stwosza 63, 80-308 Gdańsk, Poland}

\date{UNAM-CCM-2023-1\\ 10 October 2023\\ 31 January 2024 (v2)}

\maketitle

\vspace{\stretch{1}}

\begin{abstract}


We demonstrate that the recently introduced evanescent particles of a massive scalar field can be emitted and absorbed by an Unruh-DeWitt detector. In doing so the particles carry away from or deposit on the detector a quantized amount of energy, in a manner quite analogous to ordinary propagating particles. In contradistinction to propagating particles the amount of energy is less than the mass of the field, but still positive. We develop relevant methods and provide a study of the detector emission spectrum, emission probability and absorption probability involving both propagating and evanescent particles.
\end{abstract}

\vspace{\stretch{1}}
\end{titlepage}



\section{Introduction}

\emph{Evanescent particles} arise from the quantization of evanescent degrees of freedom of the field \cite{CoOe:evanescent}. Standard treatments of quantum field theory are aimed at an asymptotic description of scattering theory, where evanescent degrees of freedom are absent due to their exponential decay with distance. Thus, it is only when describing interactions from a finite distance that evanescent particles explicitly come into play \cite{Oe:quanthcyl}. Technically, they arise when quantizing the field on timelike rather than spacelike hypersurfaces. A mathematically satisfactory treatment of evanescent particles has been possible thanks to the development of the novel \emph{twisted Kähler quantization} scheme \cite{CoOe:locgenvac}. In particular, the evanescent particles have a Fock space representation with creation and annihilation operators, just like their standard propagating counterparts. It has been an open problem, however, to characterize their physical properties. It is clear from the outset that in a classical limit, evanescent particles cannot approximate the "billiard ball" behavior of propagating particles. For one, they do not have a notion of momentum associated to them. (Technically speaking we can assign them a "momentum" that is imaginary.) It is thus crucial to understand what their physical behavior is and in which sense it has anything to do with our intuition of a "particle".

Clearly, a defining property of a particle in a quantum theory should be its ability to mediate the quantized exchange of energy and other quantum numbers between systems. In the present work we study the interaction of evanescent particles with an \emph{Unruh-DeWitt (UDW) detector}. This is a pointlike quantum mechanical system interacting with a quantum field. It was first introduced by Unruh \cite{Unr:bhevap} and refined later by DeWitt \cite{dew:qgsynth} to gain a better understanding of the notions of particle and vacuum in curved spacetime, in line with the dictum that "particles are what a particle detector detects" \cite{Unruh}. In the early literature the focus was on studying the particle concept for a detector in motion \cite{BiDa:qftcurved}. In more recent years UDW models have become a powerful tool in the field of relativistic quantum information \cite{huRelativisticQuantumInformation2012} and have been treated in several situations involving the investigation of the behavior of quantum fields when interacting locally, e.g.\ \cite{brownDetectorsProbingRelativistic2013,Polo-Gomez:2022aa}. 

In this paper, we consider a pointlike UDW detector at rest in Minkowski space with a Gaussian switching function linearly coupled to a massive scalar field. To study the interaction of the detector with different particle content of the surrounding quantum field we first provide a clean description of the usual asymptotic \emph{temporal} (S-matrix) \emph{picture}. That is, we have initial particle states at asymptotically early times and final particle states at asymptotically late times. This implies the interaction picture, where the detector is switched on at intermediate times only, via the \emph{switching function}. As was shown by two of the authors, the asymptotic temporal picture is equivalent to the asymptotic \emph{radial picture} \cite{CoOe:spsmatrix,CoOe:smatrixgbf}. In the latter we have states with both incoming and outgoing particles on the asymptotic timelike hypercylinder centered at the origin, that is the celestial sphere extended over all of time. Here, we are interested instead in the hypercylinder arising from extending the sphere of \emph{finite radius}, with the UDW detector at the origin in the interior. At finite radius \emph{evanescent particles} occur in addition to propagating particles \cite{Oe:quanthcyl}. The study of their interaction with the detector, and for comparison that of propagating particles with the detector is the main objective of the present work.

In Section~\ref{sec:udwhypcyl} we recall the basics of the UDW detector and its interaction with a scalar quantum field, first in a Hamiltonian, then in a path-integral setup. We introduce the temporal (S-matrix) picture as well as the radial picture of the interaction and recall the relation between the two pictures. Section~\ref{sec:quantplane} is a brief review of the quantization of the massive Klein-Gordon field on an equal-time hyperplane, using radial coordinates. The quantization is performed in Section~\ref{sec:quanthypcyl} on the timelike hypercylinder at finite radius. In this case, evanescent modes appear in addition to the propagating ones, requiring a twisted Kähler quantization. Also, the notion of incoming and outgoing modes is established, both for the propagating and for the evanescent sector. The precise relation between the temporal and radial pictures in terms of the Hilbert spaces obtained in the previous two sections is established in Section~\ref{sec:equivalence}. In Section~\ref{sec:feynrenorm} we recall relevant amplitude and correlator formulas, establish the Feynman diagrams and Feynman rules for the UDW detector, and introduce our renormalization procedure. A general discussion of emission and absorption of both propagating and evanescent particles by the UDW detector is provided in Section~\ref{sec:emabs}. The emission spectrum is studied in some detail in Section~\ref{sec:espectrum}. The spontaneous emission probability of an excited detector is the subject of Section~\ref{sec:eprob}. Section~\ref{sec:aprob} deals with the absorption probability of a detector initially in the ground state. Extensive discussion and some outlook is provided in Section~\ref{sec:outlook}. Three appendices complement formulas and calculations for Sections~\ref{sec:udwhypcyl}, \ref{sec:quanthypcyl} and \ref{sec:aprob}. We mention in particular Appendix~\ref{sec:regularity}, which establishes necessary conditions and sufficient conditions in terms of differentiability and integrability of the switching function in order to obtain well-defined probabilities.


\section{Unruh-DeWitt detector in a timelike hypercylinder}
\label{sec:udwhypcyl}

\subsection{UDW detector in a Klein-Gordon field}

We recall the standard description of an UDW detector interacting with a real scalar field in the Hamiltonian formalism. Thus, the UDW detector is a nonrelativistic system with two states, the ground state $|g\rangle$ and the excited state $|e\rangle$. We consider a raising and a lowering operator as follows,
\begin{equation}
    \rop|g\rangle=|e\rangle,\quad \rop|e\rangle=0,\quad \lop|g\rangle=0,\quad \lop|e\rangle=|g\rangle .
\end{equation}
Taking the excitation energy to be given by $\Omega$, the free Hamiltonian of the detector may be taken to be,
\begin{equation}
    H_0=\frac{\Omega}{2}\left(\rop\lop-\lop\rop\right) .
    \label{eq:udwfreeham}
\end{equation}
The monopole interaction of the detector with a real free Klein-Gordon field is described by the interaction Hamiltonian,
\begin{equation}
    H_I(\tau)=\lambda \chi(\tau) \left(\lop+\rop\right) \hat{\phi}(x(\tau)) .
    \label{eq:udwintham}
\end{equation}
Here, $\tau$ is the proper time in the detector frame of reference, $x$ denotes the trajectory of the detector, $\hat{\phi}$ is the field operator, and $\lambda$ is a coupling constant. The \emph{switching function} $\chi$ implements an adiabatic switching on and off of the detector at early and at late times so as to ensure that we can have well-defined asymptotic initial and final states of the detector. That is, $0\le\chi(\tau)\le 1$ and $\chi(\tau)$ converges to $0$ at early and at late time. It turns out, moreover, that for mathematically well-behaved expressions we need $\chi$ to satisfy additional regularity conditions. A sufficient condition is that $\chi$ is twice differentiable and that $\chi$, $\chi'$ and $\chi''$ are all integrable, see Appendix~\ref{sec:regularity}. Our choice of $\chi$ will satisfy these conditions.

We switch to the interaction picture, where the interaction Hamiltonian takes the form,
\begin{equation}
    \tilde{H}_I(\tau)=\lambda \chi(\tau) \left(e^{-\im\Omega\tau}\lop+e^{\im\Omega\tau}\rop\right) \hat{\phi}(x(\tau)) .
\end{equation}
We denote the time-evolution operator from time $\tau_0$ to time $\tau$ by $\tilde{U}(\tau,\tau_0)$. Using time-ordered perturbation theory, this can be expanded in the Dyson series,
\begin{equation}
    \tilde{U}(\tau,\tau_0)=\sum_{n=0}^\infty (-\im)^n \int_{\tau_0}^{\tau}\xd \tau_1 \int_{\tau_0}^{\tau_1}\xd \tau_2 \cdots \int_{\tau_0}^{\tau_{n-1}}\xd \tau_n\,
    \tilde{H}_I(\tau_1) \cdots \tilde{H}_I(\tau_n) .
\end{equation}
We are interested in the transition amplitude between initial and final states of the field and the detector at asymptotically early and late times. Denote the corresponding time evolution operator by $\mathcal{S}\defeq \tilde{U}(\infty,-\infty)$. Denote initial and final states of the detector by $\psi$ and those of the field by $\Psi$. As is easy to work out,
\begin{multline}
    \langle \psi_{\Lfin}\tens\Psi_{\Lfin}, \mathcal{S}\,\psi_{\Lini}\tens\Psi_{\Lini} \rangle
    =\sum_{n=0}^\infty (-\im)^n \lambda^n \int_{-\infty}^{\infty}\xd \tau_1 \int_{-\infty}^{\tau_1}\xd \tau_2 \cdots \int_{-\infty}^{\tau_{n-1}}\xd \tau_n\\
    \chi(\tau_1)\cdots\chi(\tau_n) f_{\psi_{\Lini}\to\psi_{\Lfin}}(\tau_1,\ldots,\tau_n)\,
    \langle \Psi_{\Lfin},\hat{\phi}(x(\tau_1))\cdots\hat{\phi}(x(\tau_n))\Psi_{\Lini}\rangle .
    \label{eq:ampl}
\end{multline}
Here, the function $f$ depends on the initial and final states of the detector as follows:
\begin{align}
    f_{g\to g}(\tau_1,\ldots,\tau_n) & =e^{-\im\Omega (\tau_1-\tau_2+\tau_3\cdots -\tau_n)} \qquad n\; \text{even} ,\nonumber\\
    f_{e\to e}(\tau_1,\ldots,\tau_n) & =e^{-\im\Omega (-\tau_1+\tau_2-\tau_3\cdots +\tau_n)} \qquad n\; \text{even} ,\nonumber\\
    f_{g\to e}(\tau_1,\ldots,\tau_n) & =e^{-\im\Omega (-\tau_1+\tau_2-\tau_3\cdots -\tau_n)} \qquad n\; \text{odd} ,\nonumber\\
    f_{e\to g}(\tau_1,\ldots,\tau_n) & =e^{-\im\Omega (\tau_1-\tau_2+\tau_3\cdots +\tau_n)} \qquad n\; \text{odd} .
    \label{eq:omegafact}
\end{align}
Note that the amplitude vanishes if the detector state is unchanged and $n$ is odd or when the detector state is changed and $n$ is even.

\subsection{Path integral and detector observable}
\label{sec:piobs}

We note that the $n$-point function for the field appearing in the amplitude (\ref{eq:ampl}) is time-ordered by construction. Thus, we can obtain it from a path integral for the field. This is inspired by the path-integral treatment in \cite{BPT:pathintUDW}. We provisionally return to a setting were initial and final states live at finite times $t_1$ and $t_2$. Then,
\begin{equation}
    \langle \psi_{\Lfin}\tens\Psi_{\Lfin}, \tilde{U}(t_1,t_2)\,\psi_{\Lini}\tens\Psi_{\Lini} \rangle = \int \xD\phi\, \Psi_{\Lini}(\phi_1)\overline{\Psi_{\Lfin}(\phi_2)}\, O_{\psi_{\Lini}\to\psi_{\Lfin}}(\phi)\,     e^{\im S(\phi)} .
    \label{eq:amplpath}
\end{equation}
Here the path integral is over all field configurations between the initial time $t_1$ and the final time $t_2$. $\Psi_{\Lini}(\phi_1)$ is the Schrödinger wave function of the initial state $\Psi_{\Lini}$ evaluated on  $\phi_1$, the field configuration at time $t_1$. A corresponding statement holds for the final state. Supposing that the switching function $\chi$ vanishes outside of the time interval $[t_1,t_2]$, we can write the observable $O_{\psi_{\Lini}\to\psi_{\Lfin}}$ as follows:
\begin{multline}
    O_{\psi_{\Lini}\to\psi_{\Lfin}}(\phi)=\sum_{n=0}^\infty (-\im)^n \lambda^n \int_{-\infty}^{\infty}\xd \tau_1 \int_{-\infty}^{\tau_1}\xd \tau_2 \cdots \int_{-\infty}^{\tau_{n-1}}\xd \tau_n\\
    \chi(\tau_1)\cdots\chi(\tau_n) f_{\psi_{\Lini}\to\psi_{\Lfin}}(\tau_1,\ldots,\tau_n)\, \phi(x(\tau_1))\cdots\phi(x(\tau_n)) .
    \label{eq:detobs}
\end{multline}
We return to the asymptotic setting and introduce the following notation, where the right-hand side stands for the limit of the path integral (\ref{eq:amplpath}) for $\tau_1\to -\infty$ and $\tau_2\to\infty$,
\begin{equation}
\langle \psi_{\Lfin}\tens\Psi_{\Lfin}, \mathcal{S}\,\psi_{\Lini}\tens\Psi_{\Lini} \rangle
=\rho[O_{\psi_{\Lini}\to\psi_{\Lfin}}](\Psi_{\Lini}\tens\Psi_{\Lfin}^*) .
\end{equation}
Note a subtlety of our notation: $\Psi_{\Lfin}^*$ denotes the dual of the final state $\Psi_{\Lfin}$ in the dual Hilbert space. This reflects the fact that its wave function is complex conjugated in the path integral expression (\ref{eq:amplpath}). More formally, $\rho[O_{\psi_{\Lini}\to\psi_{\Lfin}}]$ is a map $\cH_{\Lini}\tens\cH_{\Lfin}^*\to\C$.

\subsection{Evanescent particles and the timelike hypercylinder}
\label{sec:evanescent}

In the present work we are interested in the interaction of the detector with different particle states of the Klein-Gordon field. In particular, we investigate the interaction with \emph{evanescent particles} \cite{CoOe:evanescent}. The evanescent field decays exponentially away from its source. Correspondingly, evanescent particles are detectable only at finite distance from the source.

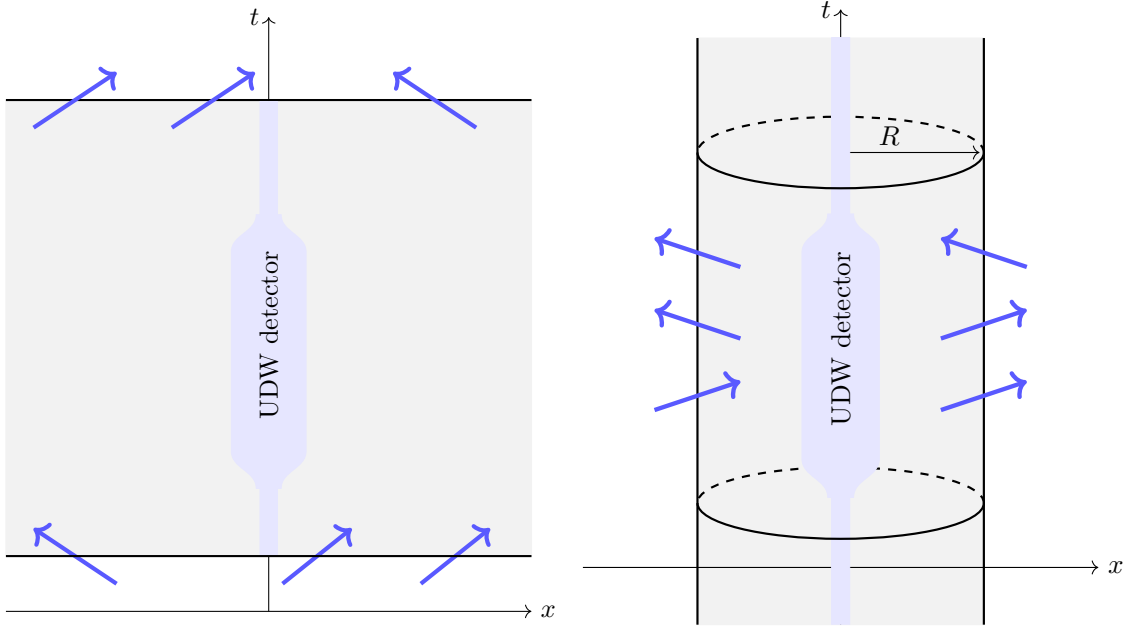
\begin{figure}
    \centering
    \resizebox{0.5\textwidth}{!}{\begin{tikzpicture}[scale=1.85]
\filldraw[color=gray!10] (-1.9,-0.4) rectangle (1.9,2.9);
\draw[->] (-1.9,-0.8) -- (1.9,-0.8) node [right] {$x$};
\draw[->] (0,-0.8) -- (0,3.5) node [left] {$t$};
\draw[thick] (-1.9,2.9) -- (1.9,2.9);
\draw[fill=black] (0,2.5) circle (0.025 cm);
\draw[ultra thick,color=blue!65,->] (-1.7,2.7) -- (-1.1,3.1);
\draw[ultra thick,color=blue!65,->] (-0.7,2.7) -- (-0.1,3.1);
\draw[ultra thick,color=blue!65,->] (1.5,2.7) -- (0.9,3.1);
\draw[ultra thick,color=blue!65,->] (-1.1,-0.6) -- (-1.7,-0.2);
\draw[ultra thick,color=blue!65,->] (0.1,-0.6) -- (0.6,-0.2);
\draw[ultra thick,color=blue!65,->] (1.1,-0.6) -- (1.6,-0.2);
    \draw[blue!10, fill=blue!10,scale=0.9] (0.07,-0.45) -- (0.07,3.21) -- (-0.07,3.21) -- (-0.07,-0.45);
    \draw[blue!10,fill=blue!10,scale=0.9] (0.1,0.1) to[out=90,in=270] (0.3,0.4) -- (0.3,2) to[out=90,in=270] (0.1,2.3) --  (-0.1,2.3) --  (-0.1,2.3) to [out=270,in=90] (-0.3,2) -- (-0.3,0.4) to [out=270,in=90]  (-0.1,0.1)--cycle;  
    \node[rotate=90,scale=1] at (0,1.2) {UDW detector};
\draw[thick] (-1.9,-0.4) -- (1.9,-0.4);

\end{tikzpicture}}%
    \resizebox{0.5\textwidth}{!}{\
\begin{tikzpicture}[scale=1.85]
\filldraw[color=gray!10] (-1,-0.8) rectangle (1,3.3);
\draw[->] (-1.8,-0.4) -- (1.8,-0.4) node [right] {$x$};
\draw[->] (0,-0.8) -- (0,3.5) node [left] {$t$};
\draw[thick] (-1,-0.8) -- (-1,3.3);
\draw[thick] (1,-0.8) -- (1,3.3);
\draw[fill=black] (0,2.5) circle (0.025 cm);
\draw[->] (0,2.5) -- (0.97,2.5);
\node at (0.34,2.62) {$R$};
\draw[dashed,thick,rotate=0] (1,2.5) arc (0:180:1 and 0.25);
\draw[dashed,thick,rotate=0] (1,0.05) arc (0:180:1 and 0.25);
\draw[ultra thick,color=blue!65,->] (0.7,0.7) -- (1.3,0.9);
\draw[ultra thick,color=blue!65,->] (0.7,1.2) -- (1.3,1.4);
\draw[ultra thick,color=blue!65,->] (-0.7,1.7) -- (-1.3,1.9);
\draw[ultra thick,color=blue!65,->] (1.3,1.7) -- (0.7,1.9);
\draw[ultra thick,color=blue!65,->] (-0.7,1.2) -- (-1.3,1.4);
\draw[ultra thick,color=blue!65,->] (-1.3,0.7) -- (-0.7,0.9);
    \draw[blue!10, fill=blue!10,scale=0.9] (0.07,-0.89) -- (0.07,3.67) -- (-0.07,3.67) -- (-0.07,-0.89);
    \draw[blue!10,fill=blue!10,scale=0.9] (0.1,0.1) to[out=90,in=270] (0.3,0.4) -- (0.3,2) to[out=90,in=270] (0.1,2.3) --  (-0.1,2.3) --  (-0.1,2.3) to [out=270,in=90] (-0.3,2) -- (-0.3,0.4) to [out=270,in=90]  (-0.1,0.1)--cycle;      
    \node[rotate=90,scale=1] at (0,1.2) {UDW detector};
\draw[thick,rotate=0] (1,0.05) arc (0:-180:1 and 0.25);
\draw[thick,rotate=0] (1,2.5) arc (0:-180:1 and 0.25);
\end{tikzpicture}}
    \caption{Temporal picture (left-hand side) vs.\ radial picture (right-hand side). In the temporal picture states of the field and the detector are fixed at initial and at final time. In the radial picture the state of the field is fixed at fixed radius, but at all time. The state of the detector remains determined at (asymptotic) initial and final time.}
    \label{fig:tempvsrad}
\end{figure}

Up to now we have followed the traditional picture of the transition amplitude between states at an initial and final time, but in all of space, see Figure~\ref{fig:tempvsrad}, left-hand side. Instead, we consider now the amplitude for states at a finite distance $R$ from the detector, but at all  times, see Figure~\ref{fig:tempvsrad}, right-hand side. In the usual temporal picture interactions may happen in all of space, but are confined between an initial time $t_1$ and a final time $t_2$. The initial and final state spaces are associated to hypersurfaces that span all of space at fixed times $t_1$, $t_2$. In contrast, in the radial picture, interactions may happen at all times, but are confined inside a fixed radius $R$. There is only one state space, associated to the hypersurface given by the sphere of radius $R$ in space, extended over all of time. This scenario and its probabilistic interpretation, was first described in \cite{Oe:kgtl}. Here, the states encode both particles that enter and that leave the interaction region. In the traditional picture, when initial and final times are taken to infinity, we recover the S-matrix. In the radial picture, when the radius is taken to infinity, we obtain an asymptotic amplitude that can be shown to be equivalent to the S-matrix \cite{CoOe:spsmatrix,CoOe:smatrixgbf}.

The state space $\cH$ at radius $R$ decomposes as a tensor product $\cH=\cH^{\Lp}\tens\cH^{\Le}$, into a \emph{propagating sector} $\cH^{\Lp}$ and an \emph{evanescent sector} $\cH^{\Le}$ \cite{Oe:quanthcyl}. In turn, the propagating sector decomposes into an \emph{incoming} and an \emph{outgoing} sector $\cH^{\Lp}=\cH^{\Lp}_{\Lin}\tens\cH^{\Lp}_{\Lout}$.\footnote{In Section~\ref{sec:quanthypcyl} we shall see that the evanescent sector also decomposes into an incoming and outgoing sector, but this is unimportant here.} There are natural identifications $\cH_{\Lini}=\cH^{\Lp}_{\Lin}$ and $\cH_{\Lfin}^*=\cH^{\Lp}_{\Lout}$ between Hilbert spaces of the temporal picture and corresponding Hilbert spaces of the radial picture which lead to an \emph{exact} equality of amplitudes between the two pictures in the asymptotic limit. Moreover, in this limit the evanescent sector $\cH^{\Le}$ no longer contributes, leading to the mentioned equivalence between the asymptotic radial amplitude and the S-matrix. On the other hand, at finite radius $R$ we still have an equality of the radial amplitude to the S-matrix if we fix the evanescent sector $\cH^{\Le}$ to the vacuum state and under the condition that any interaction or source is confined to the interior of the sphere of radius $R$ in space. In the present work this is the setting of interest.

Crucially, the relation (\ref{eq:amplpath}) between amplitude and path integral carries over to the radial setting, with exactly the same expression for the observable (\ref{eq:detobs}). For this amplitude we shall use the notation,
\begin{equation}
    \rho[O_{\psi_{\Lini}\to\psi_{\Lfin}}](\Psi)\defeq
    \int \xD\phi\, \Psi(\phi_R)\, O_{\psi_{\Lini}\to\psi_{\Lfin}}(\phi)\,     e^{\im S(\phi)} .
    \label{eq:ampludw}
\end{equation}
Here, the integral is over field configurations inside the hypercylinder $\R\times S^2_R$ formed by the sphere of radius $R$ in space, extended over all of time. $\Psi(\phi_R)$ denotes the Schrödinger wave function of the state $\Psi$, evaluated on the field configuration $\phi_R$ at radius $R$.\footnote{There are problems with the Schrödinger representation of states in $\cH_R$. However, this need not concern us as the path integral serves here really as a placeholder for the rigorous algebraic definition of the amplitude \cite{Oe:holomorphic,Oe:feynobs,CoOe:locgenvac}.} $\Psi$ is a state in the radial Hilbert space $\cH$. Note that there is no problem in combining the radial picture for the field with the temporally asymptotic picture for the detector, since the detector always remains inside the sphere of radius $R$.

In light of the equality of amplitudes it is no coincidence that we use the same notation $\rho[O_{\psi_{\Lini}\to\psi_{\Lfin}}]$ both for the temporal and the radial picture. More precisely, let $\Psi_{\Lini}\in\cH_{\Lini}$ and $\Psi_{\Lfin}\in\cH_{\Lfin}$ be initial and final states of the field in the temporal picture. Let $\Psi^{\Lp}_{\Lin}=\Psi_{\Lini}$, $\Psi^{\Lp}_{\Lout}=\Psi_{\Lfin}^*$ in $\cH^{\Lp}_{\Lin}$ and $\cH^{\Lp}_{\Lout}$ of the radial picture under the identification of Hilbert spaces. Then,
\begin{equation}
    \rho[O_{\psi_{\Lini}\to\psi_{\Lfin}}](\Psi_{\Lini}\tens\Psi_{\Lfin}^*)
    =\rho[O_{\psi_{\Lini}\to\psi_{\Lfin}}](\Psi^{\Lp}_{\Lin}\tens\Psi^{\Lp}_{\Lout}\tens\Psi^{\Le}_{\Lvac}) .
\end{equation}
Here, $\Psi^{\Le}_{\Lvac}$ denotes the vacuum state in the evanescent sector $\cH^{\Le}$.


\section{Particles on the equal-time hyperplane}
\label{sec:quantplane}

In the present section we describe the massive Klein-Gordon field and its quantization on an equal-time hypersurface in Minkowski space. The quantization is completely standard, except for the fact that we shall use radial coordinates. Our coordinates are $(t,r,\theta,\phi)$, with $\sangle$ a collective notation for angular coordinates $(\theta,\phi)$.

\subsection{Classical solutions}

We parametrize the space $L^{\C}$ of complexified solutions of the massive Klein-Gordon equation in a neighborhood of the equal-time hyperplane at time $t_0$ as follows:
\begin{equation}
 \phi(t,r,\sangle)=\int_{m}^{\infty}\xd E\, \frac{p}{2\pi} \sum_{\ls,\ms}
 \left(\phi_{E,\ls,\ms} j_{\ls}(p r) e^{-\im E t} Y_{\ls}^{\ms}(\sangle)
 + \overline{\phi}_{E,\ls,\ms} j_{\ls}(pr) e^{\im E t} Y_{\ls}^{-\ms}(\sangle)\right) .
\label{eq:tparam}
\end{equation}
Here $Y_{\ls}^{\ms}$ denote the spherical harmonics and $p\defeq\sqrt{E^2-m^2}$. Also, $j_{\ls}$ denote the spherical Bessel functions of the first kind. Real solutions satisfy $\overline{\phi}_{E,\ls,\ms}=\overline{\phi_{E,\ls,\ms}}$.\footnote{Note that the overline on the left-hand side is notation, while on the right-hand side it indicates complex conjugation.}
An important ingredient for the description of the classical dynamics and its quantization is the \emph{symplectic form} on $L$,
\begin{equation}
  \omega(\phi,\xi) =\frac{1}{2}\int \xd\sangle\,\xd r\, r^2\,
 \left(\xi(t_0,r,\sangle) \partial_t \phi(t_0,r,\sangle)- \phi(t_0,r,\sangle) \partial_t \xi(t_0,r,\sangle)\right) .
\end{equation}
In the present parametrization we obtain
\begin{equation}
  \omega(\phi,\xi)
  = -\int_{m}^\infty\xd E\frac{\im p}{8\pi}\sum_{\ls,\ms}
  \left(\overline{\xi}_{E,\ls,\ms} \phi_{E,\ls,\ms}
  -\xi_{E,\ls,\ms} \overline{\phi}_{E,\ls,\ms}\right) .
\end{equation}
Recall that the sign of the symplectic form depends on the orientation of the hypersurface. Here it corresponds to an initial hypersurface. Thus, for a final hypersurface there appears a relative minus sign.

\subsection{Quantization}

For the quantization we need to choose a vacuum. For an initial hypersurface the standard choice in our conventions is the Lagrangian subspace
\begin{equation}
    L^+=\{\phi\in L^{\C}: \phi_{E,l,m}=0\}
\end{equation}
corresponding to the negative energy modes. For a final hypersurface the standard choice is the complex conjugated Lagrangian subspace of positive energy modes,
\begin{equation}
    L^-=\overline{L^+}=\{\phi\in L^{\C}: \overline{\phi}_{E,l,m}=0\} .
\end{equation}
The decomposition $L^{\C}=L^+\oplus L^-$ written as $\phi=\phi^+ + \phi^-$ defines the standard Kähler polarization and complex structure. The induced positive-definite complex inner product on $L$ is given by
\begin{equation}
    \{\phi,\xi\}=4\im\omega(\phi^-,\xi^+)
    =\int_{m}^\infty\xd E\frac{p}{2\pi}\sum_{\ls,\ms}
    \left(\overline{\xi}_{E,\ls,\ms}\phi_{E,\ls,\ms}\right) .
    \label{eq:tbilin}
\end{equation}
This defines the commutation relations between creation and annihilation operators, which are, labeled by elements $\phi,\eta\in L$,
\begin{equation}
  [a_\eta,a_\phi^\dagger]=\{\phi,\eta\} .
  \label{eq:tcom}
\end{equation}

\subsection{Particle states}
\label{sec:tparticles}

We consider particle states on the hypercylinder, characterized in terms of energy and angular momentum quantum numbers. This is in contrast to the usual parametrization for QFT in Minkowski space, where the most convenient characterization of particle states is in terms of 3-momenta. Consider the field modes $\Phi^{E,\ls,\ms}\in L$ determined in terms of their expansion (\ref{eq:tparam}) as follows:
\begin{equation}
    (\Phi^{E,\ls,\ms})_{E',\ls',\ms'}
    = \overline{(\Phi^{E,\ls,\ms})}_{E',\ls',\ms'}
    =\sqrt{\frac{2\pi}{p}}\delta_{\ls,\ls'}\delta_{\ms,\ms'}\delta(E-E') .
    \label{eq:partt}
\end{equation}
The creation and annihilation operators satisfy the commutation relations, due to (\ref{eq:tbilin}) and (\ref{eq:tcom}),
\begin{equation}
    [a_{E,\ls,\ms},a^\dagger_{E',\ls',\ms'}]=\delta_{\ls,\ls'}\delta_{\ms,\ms'} \delta(E-E') . \label{eq:ccr}
\end{equation}


\section{Particles on the timelike hypercylinder}
\label{sec:quanthypcyl}

In the present section we describe the massive Klein-Gordon field in radial coordinates and its quantization on the \emph{timelike hypercylinder} $\R\times S^2_R$, where $S^2_R$ denotes the two-sphere in space of radius $R$, centered at the origin. We largely give a summary of the treatment of \cite{Oe:quanthcyl}, although with some modifications. In particular, we treat propagating and evanescent modes in a more uniform way here, more in line with the work \cite{Oe:holomorphic}.

\subsection{Classical solutions}

We parametrize the space $L^{\C}$ of complexified solutions of the massive Klein-Gordon equation in a neighborhood of the hypercylinder $\R\times S^2_R$ as follows:
\begin{multline}
 \phi(t,r,\sangle)=\int_{0}^{\infty}\xd E\, \frac{p}{4\pi} \sum_{\ls,\ms}
 \left(\left(\phi_{E,\ls,\ms}^\Lout d_{\ls}(p r) + \phi_{E,\ls,\ms}^\Lin \overline{d_{\ls}(pr)}\right) e^{-\im E t} Y_{\ls}^{\ms}(\sangle) \right.\\
 \left. + \left(\phi_{E,\ls,\ms}^\Lcout \overline{d_{\ls}(p r)} + \phi_{E,\ls,\ms}^\Lcin d_{\ls}(pr)\right) e^{\im E t} Y_{\ls}^{-\ms}(\sangle)\right) .
\label{eq:pparam}
\end{multline}
Here $Y_{\ls}^{\ms}$ denote the spherical harmonics and $p\defeq\sqrt{|E^2-m^2|}$. Also,
\begin{equation}
  d_\ls(pr)\defeq\begin{cases} j_{\ls}(p r)+\im n_{\ls}(p r) &
    \text{if}\; E>m\\
    \im^{-\ls} j_{\ls}(\im p r)-\im^{\ls} n_{\ls}(\im p r) &
    \text{if}\; E<m \end{cases},
    \label{eq:besseld}
\end{equation}
where $j_{\ls}$ and $n_{\ls}$ are the spherical Bessel functions of the first and second kind, respectively. Note that the linear combination $j_{\ls}+\im n_{\ls}$ is a spherical Bessel function of the third kind. We remark that $\im^{-\ls} j_{\ls}(\im p r)$ is real while $\im^{\ls} n_{\ls}(\im p r)$ is imaginary. Real solutions satisfy $\phi_{E,\ls,\ms}^\Lcout=\overline{\phi_{E,\ls,\ms}^\Lout}$ and $\phi_{E,\ls,\ms}^\Lcin=\overline{\phi_{E,\ls,\ms}^\Lin}$.\footnote{Note that the overlines on the left-hand sides are notations, while on the right-hand sides they indicate complex conjugation.}

Solutions that are regular in the interior $M$ of the hypercylinder are those where only Bessel functions of the first kind (and its analytical continuations), but not of the second kind contribute. We denote the subspace of these solutions by $L_M^{\C}\subseteq L^{\C}$,
\begin{equation}
  L_M^{\C}
  =\{\phi\in L^{\C} : \phi_{E,\ls,\ms}^\Lin=\phi_{E,\ls,\ms}^\Lout,\; \phi_{E,\ls,\ms}^\Lcin=\phi_{E,\ls,\ms}^\Lcout \} .
  \label{eq:propint}
\end{equation}
We also note that $k_{\ls}(z)=(\pi/2)(-\im^{\ls} j_{\ls}(\im z)-\im^{\ls+1} n_{\ls}(\im z))$ and $\tilde{k}_{\ls}(z)=k_{\ls}(-z)$ are modified spherical Bessel functions that are real for $z\in\R$. $k_{\ls}(z)$ decays exponentially for increasing $z$. As long as $z$ is not small, $\tilde{k}_{\ls}(z)$ decays exponentially for decreasing $z$.

An important ingredient for the description of the classical dynamics and its quantization is the \emph{symplectic form} on $L$,\footnote{Compared with \cite{Oe:quanthcyl}, we set $\omega=\omega_{\oR}=-\omega_R$.}
\begin{equation}
  \omega(\phi,\xi) =\frac{R^2}{2}\int \xd t\,\xd\sangle\,
 \left(\xi(t,R,\sangle) \partial_r \phi(t,R,\sangle)- \phi(t,R,\sangle) \partial_r \xi(t,R,\sangle)\right) .
\end{equation}
In the present parametrization we obtain
\begin{equation}
  \omega(\phi,\xi)
  = -\int_{0}^\infty\xd E\frac{\im p}{8\pi}\sum_{\ls,\ms}
  \left(\xi_{E,\ls,\ms}^\Lout \phi_{E,\ls,\ms}^\Lcout
  +\xi_{E,\ls,\ms}^\Lcin \phi_{E,\ls,\ms}^\Lin
  -\xi_{E,\ls,\ms}^\Lcout \phi_{E,\ls,\ms}^\Lout
  -\xi_{E,\ls,\ms}^\Lin \phi_{E,\ls,\ms}^\Lcin\right) .
\end{equation}

\subsection{Incoming and outgoing modes}
\label{sec:inout}

In expression (\ref{eq:besseld}) we can appreciate the distinction between \emph{propagating modes} (with $E>m$) and \emph{evanescent modes} (with $E<m$), which we denote by $L^{\C}=L^{\Lp,\C}\oplus L^{\Le,\C}$. The latter only occur in the case where the field is massive. For a spacelike hypersurface, the latter modes would not be present even in the massive case as there are no such modes that are well-defined and bounded in all of space. The propagating modes are described in terms of Bessel functions of the third kind, which asymptotically (for large radius) behave like sine and cosine waves with an inverse radial decay of the amplitude. In contrast, the evanescent modes are described by modified spherical Bessel functions that show exponential behavior (growth or decay) in the radial direction.

For propagating modes, their asymptotic form shows that they can always be decomposed into components consisting of waves that move either radially into the origin or radially out of the origin. This decomposition into \emph{incoming} and \emph{outgoing} solutions (with respect to the interior of the hypercylinder), can be formalized as follows, as already suggested by our previous notation:
\begin{align}
  L^{\Lin,\C} & =\{\phi\in L^{\C} : \phi_{E,\ls,\ms}^\Lout=0,\;  \phi_{E,\ls,\ms}^\Lcout=0\},
  \label{eq:inmodes} \\
  L^{\Lout,\C} & =\{\phi\in L^{\C} : \phi_{E,\ls,\ms}^\Lin=0,\;  \phi_{E,\ls,\ms}^\Lcin=0\} .
  \label{eq:outmodes}
\end{align}
We have the direct sum decomposition $L^{\C}=L^{\Lin,\C}\oplus L^{\Lout,\C}$.
For evanescent modes, there is apparently no comparable sense in which they may be considered incoming or outgoing with respect to the hypercylinder since they are not oscillating in the radial direction. However, as we shall see shortly, there is a well-defined sense in which the evanescent modes can also be decomposed in terms of incoming and outgoing modes. What is more, this decomposition is precisely as indicated implicitly in the parametrization (\ref{eq:pparam}). To see this, consider the flow of energy through the timelike hypercylinder as measured by the energy-momentum tensor. More precisely, we integrate the $T_{0 i}$ component of the energy-momentum tensor over the sphere of radius $r$ in space, where $i$ denotes the spatial direction perpendicular to the sphere. We write this component as $T_{0 r}$. Additionally, we integrate over time so as to obtain the total flux $F$ through the hypercylinder. For a real solution $\phi$ we obtain,
\begin{align}
  F(\phi) & = R^2 \int \xd t\,\xd\sangle\, T_{0 r}(\phi) = R^2 \int \xd t\,\xd\sangle\, (\partial_0\phi)(t,R,\sangle) (\partial_r\phi)(t,R,\sangle) \\
  & = \im R^2 \int_{0}^\infty\xd E\frac{p^3 E}{8\pi}\sum_{\ls,\ms}
  \left(\left|\phi_{E,\ls,\ms}^\Lin\right|^2
  - \left|\phi_{E,\ls,\ms}^\Lout\right|^2\right)
  \left(d_{\ls}(p R) \overline{d_{\ls}'(p R)} - \overline{d_{\ls}(p R)} d_{\ls}'(p R)\right) .
  \nonumber \\
  & = \int_{0}^\infty\xd E\frac{p E}{4\pi}\sum_{\ls,\ms}
  \left(\left|\phi_{E,\ls,\ms}^\Lin\right|^2
  - \left|\phi_{E,\ls,\ms}^\Lout\right|^2\right) .
\end{align}
From this expression we can read off that, indeed, the incoming modes carry energy flux into the hypercylinder, while the outgoing modes carry energy flux out of the hypercylinder. With our conventions, particularly expression (\ref{eq:besseld}), this is true both for the propagating and the evanescent sector.

\subsection{Quantization in the propagating sector}
\label{sec:quantprop}

For the propagating solutions, quantization proceeds in close analogy to the standard Kähler quantization which is well established for spacelike hypersurfaces. The key ingredient is the Lagrangian subspace encoding the physical vacuum in the form of the Wick-rotated asymptotic vanishing condition of the field, here,
\begin{equation}
  L_{X}^{\Lp,\C}=\{\phi\in L^{\Lp,\C} : \phi_{E,\ls,\ms}^\Lin=0, \phi_{E,\ls,\ms}^\Lcout=0\} .
\end{equation}
The decomposition $L^{\Lp,\C}=L_{X}^{\Lp,\C} \oplus \overline{L_{X}^{\Lp,\C}}$, which we write as $\phi=\phi^+ + \phi^-$, thus defines a Kähler polarization, analogous to the decomposition into positive and negative energy solutions for spacelike hypersurfaces. We obtain a complex structure and a positive-definite complex inner product on $L^{\Lp}$,\footnote{Compared with \cite{Oe:quanthcyl} we set $\{\cdot,\cdot\}=\{\cdot,\cdot\}_{\oR}$.}
\begin{equation}
  \{\phi,\xi\}^{\Lp}=4\im\omega^{\Lp}(\phi^-,\xi^+)
  =\int_{m}^\infty\xd E\frac{p}{2\pi}\sum_{\ls,\ms}
  \left(\xi_{E,\ls,\ms}^\Lout\phi_{E,\ls,\ms}^\Lcout
  + \xi_{E,\ls,\ms}^\Lcin\phi_{E,\ls,\ms}^\Lin\right) .
  \label{eq:pbilin}
\end{equation}
The corresponding quantization defines the commutation relations between creation and annihilation operators, which are, labeled by elements $\phi,\eta\in L^{\Lp}$,
\begin{equation}
  [a_\eta,a_\phi^\dagger]=\{\phi,\eta\}^{\Lp} .
  \label{eq:pcom}
\end{equation}

\subsection{Quantization in the evanescent sector}
\label{sec:quantev}

For evanescent modes the physical vacuum in the exterior of the hypercylinder is determined by a (non-Wick-rotated) decaying boundary condition \cite{CoOe:vaclag}. This is encoded in the Lagrangian subspace given by \cite{Oe:quanthcyl}
\begin{equation}
  L_{X}^{\Le,\C}=\{\phi\in L^{\Le,\C} : \phi_{E,\ls,\ms}^{\Lout}=-(-1)^{\ls}\im \phi_{E,\ls,\ms}^{\Lin},\; \phi_{E,\ls,\ms}^{\Lcout}=(-1)^{\ls}\im \phi_{E,\ls,\ms}^{\Lcin}\} .
  \label{eq:evandec}
\end{equation}
$L_{X}^{\Le,\C}$ defines a real polarization rather than a Kähler polarization. This requires the application of the novel twisted Kähler quantization scheme developed for this purpose \cite{CoOe:locgenvac}. To this end we choose correspondingly a vacuum for the interior of the hypercylinder in terms of a complementary Lagrangian subspace. This is determined by a decay condition to the interior \cite{Oe:quanthcyl},
\begin{equation}
  L_{<R}^{\Le,\C}=\{\phi\in L^{\Le,\C} : \phi_{E,\ls,\ms}^{\Lout}=(-1)^{\ls}\im \phi_{E,\ls,\ms}^{\Lin},\; \phi_{E,\ls,\ms}^{\Lcout}=-(-1)^{\ls}\im \phi_{E,\ls,\ms}^{\Lcin}\} .
  \label{eq:evaninc}
\end{equation}
Moreover, we need a real structure $\alpha:L^{\Le,\C}\to L^{\Le,\C}$, compatible with the symplectic structure, the polarization, and positive definite \cite{CoOe:locgenvac}. It turns out that imposing spacetime symmetries and a certain compatibility condition (called interior compatibility) determines a unique real structure given in \cite{Oe:quanthcyl}, see relations \eqref{eq:alpha} of Appendix~\ref{sec:extraquant}.
Decomposing $L^{\Le,\C}=L_{X}^{\Le,\C} \oplus L_{<R}^{\Le,\C}$ with notation $\phi=\phi^+ + \phi^-$ we have for the bilinear form corresponding to (\ref{eq:pbilin}),
\begin{gather}
  \{\phi,\xi\}^{\Le}=4\im\omega^{\Le}(\phi^-,\xi^+)
  =\int_0^{m}\xd E\frac{p}{4\pi}\sum_{\ls,\ms} \nonumber\\
  \left(\xi_{E,\ls,\ms}^\Lout\phi_{E,\ls,\ms}^\Lcout
  - \xi_{E,\ls,\ms}^\Lcout\phi_{E,\ls,\ms}^\Lout
  + \xi_{E,\ls,\ms}^\Lcin\phi_{E,\ls,\ms}^\Lin
  - \xi_{E,\ls,\ms}^\Lin\phi_{E,\ls,\ms}^\Lcin\right. \nonumber\\
  \left. +(-1)^{\ls+1}\im\left(\xi_{E,\ls,\ms}^\Lin\phi_{E,\ls,\ms}^\Lcout
  + \xi_{E,\ls,\ms}^\Lcin\phi_{E,\ls,\ms}^\Lout
  + \xi_{E,\ls,\ms}^\Lout\phi_{E,\ls,\ms}^\Lcin
  + \xi_{E,\ls,\ms}^\Lcout\phi_{E,\ls,\ms}^\Lin\right)\right)
  \label{eq:ebilin}
\end{gather}
This is positive-definite on the $\alpha$-twisted real solution space $L^{\Le,\alpha}=\{\phi\in L^{\Le,\C} : \alpha(\phi)=\phi\}$.
The creation and annihilation operators are labeled by elements of $L^{\Le,\alpha}$, and satisfy the commutation relations analogous to the propagating sector (\ref{eq:pcom}),
\begin{equation}
  [a_\eta,a_\phi^\dagger]=\{\phi,\eta\}^{\Le} .
  \label{eq:ecom}
\end{equation}
The total Hilbert space of states associated to the hypercylinder is the (completed) tensor product of the Hilbert spaces for the propagating and evanescent sectors, $\cH=\cH^{\Lp}\tens\cH^{\Le}$.

For the evanescent sector, the parametrization of creation and annihilation operators in terms of elements of the twisted phase space $L^{\Le,\alpha}$ instead of the ordinary phase space $L^{\Le}$ poses a problem in terms of the semiclassical interpretation of states. To remedy this we introduce a linear mapping $I^{\Le}:L^{\Le}\to L^{\Le,\alpha}$, bringing the two spaces into correspondence \cite{CoOe:locgenvac,Oe:quanthcyl}. For our present parametrization this is provided by relations \eqref{eq:idralphaexp} of Appendix~\ref{sec:extraquant}. This appendix also contains additional expressions arising in the quantization problem, required for some calculations in later sections.

\subsection{Particle states}
\label{sec:particles}

In this section we consider particle states on the hypercylinder, characterized in terms of energy and angular-momentum quantum numbers, as well as a binary quantum number distinguishing incoming from outgoing particles (in the sense of Section~\ref{sec:inout}). Consider the field modes $\Phi^{\Lin,E,\ls,\ms},\Phi^{\Lout,E,\ls,\ms}\in L$ determined in terms of their expansion (\ref{eq:pparam}) as follows:
\begin{align}
    (\Phi^{\Lout,E,\ls,\ms})_{E',\ls',\ms'}^\Lout
    = (\Phi^{\Lout,E,\ls,\ms})_{E',\ls',\ms'}^\Lcout
    & =\sqrt{\frac{2\pi}{p}}\delta_{\ls,\ls'}\delta_{\ms,\ms'}\delta(E-E') ,
    \label{eq:partpin}  \\
    (\Phi^{\Lin,E,\ls,\ms})_{E',\ls',\ms'}^\Lin
    = (\Phi^{\Lin,E,\ls,\ms})_{E',\ls',\ms'}^\Lcin
    & =\sqrt{\frac{2\pi}{p}}\delta_{\ls,\ls'}\delta_{\ms,\ms'}\delta(E-E') .
    \label{eq:partpout}
\end{align}
The other coefficients are zero. In the propagating sector (see Section~\ref{sec:quantprop}) the creation and annihilation operators satisfy the commutation relations, due to (\ref{eq:pbilin}) and (\ref{eq:pcom}),
\begin{align}
    [a_{\Lin,E,\ls,\ms},a^\dagger_{\Lin,E',\ls',\ms'}]=\delta_{\ls,\ls'}\delta_{\ms,\ms'} \delta(E-E'), \label{eq:ccrin} \\
    [a_{\Lout,E,\ls,\ms},a^\dagger_{\Lout,E',\ls',\ms'}]=\delta_{\ls,\ls'}\delta_{\ms,\ms'} \delta(E-E') . \label{eq:ccrout}
\end{align}
Commutators involving both incoming and outgoing particles vanish.

In the evanescent sector, the modes we have defined, cannot be directly considered for quantization as they live in the real phase space $L^{\Le}$ rather than in the $\alpha$-twisted phase space $L^{\Le,\alpha}$. We use the identification map $I^{\Le}$ given by (\ref{eq:idralphaexp}) to obtain the corresponding elements in $L^{\Le,\alpha}$ and denote them with a tilde, $\tilde{\Phi}^{\bullet}=I^{\Le}(\Phi^{\bullet})$. This yields,
\begin{align}
    (\tilde{\Phi}^{\Lout,E,\ls,\ms})_{E',\ls',\ms'}^{\Lout} & = (\tilde{\Phi}^{\Lout,E,\ls,\ms})_{E',\ls',\ms'}^{\Lcout} ={\sqrt\frac{\pi}{p}}\delta_{\ls,\ls'}\delta_{\ms,\ms'}\delta(E-E') , \nonumber \\
    (\tilde{\Phi}^{\Lout,E,\ls,\ms})_{E',\ls',\ms'}^{\Lin} & = (\tilde{\Phi}^{\Lout,E,\ls,\ms})_{E',\ls',\ms'}^{\Lcin} =(-1)^{\ls}\im\sqrt{\frac{\pi}{p}}\delta_{\ls,\ls'}\delta_{\ms,\ms'}\delta(E-E') , \label{eq:parteout}\\
    (\tilde{\Phi}^{\Lin,E,\ls,\ms})_{E',\ls',\ms'}^{\Lout} & = (\tilde{\Phi}^{\Lin,E,\ls,\ms})_{E',\ls',\ms'}^{\Lcout} =(-1)^{\ls}\im\sqrt{\frac{\pi}{p}}\delta_{\ls,\ls'}\delta_{\ms,\ms'}\delta(E-E') , \nonumber \\
    (\tilde{\Phi}^{\Lin,E,\ls,\ms})_{E',\ls',\ms'}^{\Lin} & = (\tilde{\Phi}^{\Lin,E,\ls,\ms})_{E',\ls',\ms'}^{\Lcin} =\sqrt{\frac{\pi}{p}}\delta_{\ls,\ls'}\delta_{\ms,\ms'}\delta(E-E') , \label{eq:partein}
\end{align}
Using the obvious notation for the corresponding creation and annihilation operators, from (\ref{eq:ebilin}) and (\ref{eq:ecom}), these satisfy the same commutation relations (\ref{eq:ccrin}) and (\ref{eq:ccrout}), as in the propagating case.

The commutation relations (\ref{eq:ccrin}) and (\ref{eq:ccrout}), lead to a simple \emph{completeness relation} for the 1-particle subspace $\cH^1\subseteq \cH$,
\begin{equation}
    \id^1=\sum_{\ls,\ms}\int_{0}^{\infty}\xd E\,
      \left(P_{\Lin,E,\ls,\ms} + P_{\Lout,E,\ls,\ms}\right) .
      \label{eq:hs1compl}
\end{equation}
Here, $P_{\bullet,E,\ls,\ms}$ represents a projection-like operator onto the corresponding state,\footnote{The operators $P_{\bullet,E,\ls,\ms}$ are not actual projection operators due to the singular nature of the commutation relations with respect to the energy variable.}
\begin{equation}
   P_{\bullet,E,\ls,\ms} =a^\dagger_{\bullet,E,\ls,\ms}|0\rangle
    \langle 0 | a_{\bullet,E,\ls,\ms} .
\end{equation}
The completeness relation extends straightforwardly to the $n$-particle sector of the state space. In that case there will be $n$ sums and integrals over the energy and angular-momentum quantum numbers.

For the calculation of amplitudes it will be instrumental to consider the following decomposition of the complexified phase space arising from the choice of vacuum,
\begin{equation}
  L^{\C}=L_M^{\C}\oplus L_X^{\C},\qquad\text{written as}\qquad \xi=\xi^{\Lint}+\xi^{\Lext}.
  \label{eq:vacdec}
\end{equation}
We calculate the components $\xi^{\Lint}$ of the (twisted) phase space elements $\xi$ encoding the particle states. For propagating particles, $\xi=\Phi^{\bullet}\in L^{\Lp}$, we obtain with (\ref{eq:partpin}) and (\ref{eq:partpout}),
\begin{align}
    (\Phi^{\Lout,E,\ls,\ms})^{\Lint}(t,r,\sangle)
    & = \sqrt{\frac{p}{2\pi}} j_{\ls}(p r) e^{\im E t} Y_{\ls}^{-\ms}(\sangle) ,
    \label{eq:propoutint}\\
    (\Phi^{\Lin,E,\ls,\ms})^{\Lint}(t,r,\sangle)
    & = \sqrt{\frac{p}{2\pi}} j_{\ls}(p r) e^{-\im E t} Y_{\ls}^{\ms}(\sangle) .
    \label{eq:propinint}
\end{align}
For the evanescent particles, $\xi=\tilde{\Phi}^{\bullet}\in L^{\Le,\alpha}$ we obtain with (\ref{eq:parteout}) and (\ref{eq:partein}),
\begin{align}
    (\tilde{\Phi}^{\Lout,E,\ls,\ms})^{\Lint}(t,r,\sangle)
    & = \sqrt{\frac{p}{4\pi}} (1+(-1)^{\ls}\im) \im^{-\ls} j_{\ls}(\im p r) e^{\im E t} Y_{\ls}^{-\ms}(\sangle),
    \label{eq:evanoutint} \\
    (\tilde{\Phi}^{\Lin,E,\ls,\ms})^{\Lint}(t,r,\sangle)
    & = \sqrt{\frac{p}{4\pi}} (1+(-1)^{\ls}\im) \im^{-\ls} j_{\ls}(\im p r) e^{-\im E t} Y_{\ls}^{\ms}(\sangle) .
    \label{eq:evaninint}
\end{align}


\section{Equivalence of amplitudes}
\label{sec:equivalence}

In terms of the quantizations discussed in Sections~\ref{sec:quantplane} and \ref{sec:quanthypcyl}, the equivalence of amplitudes for the temporal and the radial picture mentioned in Section~\ref{sec:evanescent} can be expressed as follows.\footnote{Tools and notations used to express the equivalence here differ considerably from those used in the original papers \cite{CoOe:spsmatrix,CoOe:smatrixgbf}. The present methods are based on more recent works, in particular \cite{Oe:holomorphic,Oe:feynobs,CoOe:vaclag,CoOe:locgenvac}.} Underlying this are two maps from the space of solutions on an equal-time hyperplane to the propagating sector of the space of solutions on the timelike hypercylinder,
\begin{equation}
    \tau^{\Lini}:L\to L^{\Lp}, \quad\text{and}\quad \tau^{\Lfin}:L\to L^{\Lp} .
\end{equation}
Here, $L$ denotes the space of solutions for the equal-time hyperplane (Section~\ref{sec:quantplane}), while $L^{\Lp}$ denotes the propagating sector of the space of solutions for the timelike hypercylinder (Section~\ref{sec:quanthypcyl}). These maps are given as follows:
\begin{gather}
    (\tau^{\Lini}(\phi))^{\Lin}_{E,\ls,\ms}=\phi_{E,\ls,\ms},\quad (\tau^{\Lini}(\phi))^{\Lcin}_{E,\ls,\ms}=\overline{\phi}_{E,\ls,\ms}, \quad
    (\tau^{\Lini}(\phi))^{\Lout}_{E,\ls,\ms}=(\tau^{\Lini}(\phi))^{\Lcout}_{E,\ls,\ms}=0,\\
    (\tau^{\Lfin}(\phi))^{\Lout}_{E,\ls,\ms}=\overline{\phi}_{E,\ls,\ms},\quad (\tau^{\Lfin}(\phi))^{\Lcout}_{E,\ls,\ms}=\phi_{E,\ls,\ms}, \quad
    (\tau^{\Lfin}(\phi))^{\Lin}_{E,\ls,\ms}=(\tau^{\Lfin}(\phi))^{\Lcin}_{E,\ls,\ms}=0 .
\end{gather}
The Fock quantization of these maps gives rise to the isomorphisms of Hilbert spaces $\cH_{\Lini}\to \cH^{\Lp}_{\Lin}$ and $\cH_{\Lfin}^*\to \cH^{\Lp}_{\Lout}$. The nontrivial result of \cite{CoOe:spsmatrix,CoOe:smatrixgbf} is that this establishes an equivalence of interacting amplitudes between the temporal and the radial picture (Section~\ref{sec:evanescent}) at the level of perturbation theory.
It is useful to express this more concretely at the level of states. To this end we use the following notation for $n$-particle states, where $\xi_1,\ldots,\xi_n\in L$,
\begin{equation}
    \Psi_{\xi_1,\ldots,\xi_n}\defeq a^\dagger_{\xi_1}\cdots a^\dagger_{\xi_n} \Psi_{\Lvac} .
\end{equation}
Here, $\Psi_{\Lvac}$ denotes the vacuum state.
We may now express the equivalence of amplitudes as follows,
\begin{equation}
    \rho[F](\Psi_{\xi_1,\ldots,\xi_n}\tens\Psi_{\eta_1,\ldots,\eta_m}^*)
    =\rho[F](\Psi_{\tau^{\Lini}(\xi_1),\ldots,\tau^{\Lini}(\xi_n),\tau^{\Lfin}(\eta_1),\ldots,\tau^{\Lfin}(\eta_m)}) .
\end{equation}
The left-hand side represents the amplitude in the temporal picture, for initial $n$-particle state $\Psi_{\xi_1,\ldots,\xi_n}$ and final $m$-particle state $\Psi_{\eta_1,\ldots,\eta_m}$. The right-hand side represents the amplitude in the radial picture, for the $(n+m)$-particle state $\Psi_{\tau^{\Lini}(\xi_1),\ldots,\tau^{\Lini}(\xi_n),\tau^{\Lfin}(\eta_1),\ldots,\tau^{\Lfin}(\eta_m)}$ in the propagating sector of the Hilbert space. The symbol $F$ represents a generic observable or source, restricted to the interior of the hypercylinder and vanishing at positive and negative infinite time. As we have seen in Section~\ref{sec:udwhypcyl}, the UDW detector with fixed initial and final state induces precisely such an observable.
Note that the structure of the maps $\tau$ means that initial particles are mapped to incoming particles and final particles are mapped to outgoing particles, as one should expect. Furthermore, the particle states with definite quantum numbers defined in Sections~\ref{sec:tparticles} and \ref{sec:particles} are mapped to each other conserving the quantum numbers, again, as one should expect. That is, an initial particle with given energy and angular momentum corresponds to an incoming particle with the same energy and angular momentum. The same for final particles corresponding to outgoing particles.

In the remainder of this work, we focus exclusively on the radial picture in evaluating amplitudes and probabilities for the interaction of the UDW detector with particles. However, when restricting to the vacuum state in the evanescent sector, due to the discussed equivalence, this implies the exact corresponding results for the temporal picture. We shall comment on this relation from time to time.


\section{Amplitudes, Feynman diagrams and renormalization}
\label{sec:feynrenorm}

In the present section we show how to evaluate the amplitudes (\ref{eq:ampludw}) describing the interaction of the UDW detector with states of the Klein-Gordon field.

\subsection{Amplitudes}

We review some (generalizations of) mostly well-known identities from quantum field theory. As a specific reference adapted to the present framework we point the reader to \cite{CoOe:locgenvac}.
We write $\rho(\Psi)$ to denote the free field theory amplitude for a state $\Psi\in\cH$ on the hypercylinder. As noted previously, we can think of this amplitude as given by the Feynman path integral over field configurations in the interior of the hypercylinder.
We recall that the free amplitude for an $n$-particle state decomposes into
products of 2-particle amplitudes if $n=2m$ is even, and vanishes otherwise,
\begin{equation}
    \rho(\Psi_{\xi_1,\ldots,\xi_n})
    =\frac{1}{2^m\, m!}\sum_{\sigma\in S^{2m}}
    \prod_{j=1}^m \rho(\Psi_{\xi_{\sigma(2j-1)},\xi_{\sigma(2j)}}) .
    \label{eq:freepairing}
\end{equation}
Here, $\sigma$ runs over the elements of the permutation group $S^{2m}$ of $2m$ elements.
The 2-particle amplitude can be written as the following bilinear symmetric expression,
\begin{equation}
    \rho(\Psi_{\xi_1,\xi_2}) =\{\xi_1,\xi_2^{\Lint}\} =\{\xi_2,\xi_1^{\Lint}\}
    =\{\xi_1,u(\xi_2)\}=\{\xi_2,u(\xi_1)\} .
    \label{eq:2ampl}
\end{equation}
Apart from the structures defined in Section~\ref{sec:quanthypcyl} we are also using here the map $u$ defined in Appendix~\ref{sec:extraquant}.

Before considering the amplitudes arising in the field theory from the interaction with the UDW detector (compare Sections~\ref{sec:piobs} and \ref{sec:evanescent}), we consider amplitudes with generic insertions of observables. We use the notation $\rho[O](\Psi)$ to denote the amplitude for a state $\Psi$ with observable $O$ inserted.
We recall that it is convenient in the description of scattering processes to distinguish contributions according to whether or not external particles participate in the scattering process. This leads to the following decomposition of the interacting amplitude \cite[Section~6.6]{CoOe:locgenvac},
\begin{equation}
    \rho[O](\Psi_{\xi_1,\dots,\xi_n})
    = \sum_{m=0}^{\lfloor n/2 \rfloor}  \sum_{\sigma\in S^n} \frac{1}{(2m)!\, (n-2m)!}\, \rho\left(\Psi_{\xi_{\sigma(1)},\dots,\xi_{\sigma(2m)}}\right)
    \rho_{\text{c}}[O]\left(\Psi_{\xi_{\sigma(2m+1)},\dots,\xi_{\sigma(n)}}\right) .
    \label{eq:decon}
\end{equation}
Here, $\rho_{\text{c}}[O]$ denotes the \emph{connected amplitude}\footnote{This is not to be confused with the notion of connected Feynman diagram.}, where all external particles participate in the scattering process. In terms of Feynman diagrams, the connected amplitude only comprises diagrams where all external particles are connected to interaction vertices. For a generic Feynman diagram, the other particles are paired up as seen in expression (\ref{eq:freepairing}). Each of these pairs thus represents a single particle that enters and leaves the hypercylinder undisturbed, without interacting.

We proceed to consider the connected amplitude of an arbitrary state $\Psi$ with a product $D_1\cdots D_n$ of linear observables. This decomposes as follows \cite[Section~6.5]{CoOe:locgenvac}
\begin{multline}
    \rho_{\text{c}}[D_1 \cdots D_n](\Psi) = \\
    \sum_{m=0}^{\lfloor n/2 \rfloor} \sum_{\sigma\in S^n} \frac{1}{2^m\, m!\, (n-2m)!} \rho_{\text{c}}[\no{D_{\sigma(2m+1)}\cdots D_{\sigma(n)}}](\Psi)\, \prod_{j=1}^m\, \rho[D_{\sigma(2j-1)} D_{\sigma(2j)}](\Psi_{\Lvac}) .
    \label{eq:denorm}
\end{multline}
Here, $\rho_{\text{c}}[\no{O}](\Psi)$ denotes the connected amplitude with the observable $O$ quantized according to normal ordering. This in turn is nonvanishing only if the degree of the observable coincides with the particle number,
\begin{equation}
    \rho_{\text{c}}[\no{D_1\cdots D_n}](\Psi_{\xi_1,\ldots,\xi_n})
    =\sum_{\sigma\in S^n}\prod_{k=1}^n \rho[D_k](\Psi_{\xi_{\sigma(k)}}) .
    \label{eq:nopair}
\end{equation}
The last two relations have the following interpretation in terms of Feynman diagrams. The observable $D_1\cdots D_n$ represents a single or a product of various vertices. In total these vertices have $n$ legs. In each Feynman diagram some of these legs are connected to each other with propagators. This is represented by the rightmost term in expression (\ref{eq:denorm}), the vacuum amplitudes with pairs of linear observables. The other legs, represented by the normal ordered amplitude appearing in expression (\ref{eq:denorm}) carry the external particle lines, as becomes clear from expression (\ref{eq:nopair}).

It remains to evaluate the amplitude for a linear observable on a single-particle state. For $\xi\in L^{\Lp}\oplus L^{\Le,\alpha}$, this is,
\begin{equation}
    \rho[D](\Psi_{\xi})=\sqrt{2} D(\xi^{\Lint}) .
    \label{eq:partend}
\end{equation}
(Recall the decomposition (\ref{eq:vacdec}).)

\subsection{Feynman diagrams and Feynman rules}

We are now ready to evaluate amplitudes for the interaction of multiparticle states of the field with the UDW detector. Thus, we consider amplitudes with the observable $O_{\psi_{\Lin}\to\psi_{\Lout}}$ given by expression (\ref{eq:detobs}). With the previously established relations, we can read off Feynman rules and Feynman diagrams. We focus on the integrand at a fixed order $m$ (in $\lambda$) contribution to the observable,
\begin{equation}
    (-\im)^m \lambda^m 
    \chi(\tau_1)\cdots\chi(\tau_m) f_{\psi_{\Lin}\to\psi_{\Lout}}(\tau_1,\ldots,\tau_m)\, \phi(x(\tau_1))\cdots\phi(x(\tau_m)) .
    \label{eq:udwintegrand}
\end{equation}
This yields Feynman diagrams with $m$ vertices, one for each time variable. These Feynman diagrams naturally live in a spacetime representation, rather than the more usual momentum space representation. The vertex $k$ is located in spacetime at event $x(\tau_k)$, i.e., the event that the detector passes at proper time $\tau_k$. It is convenient to depict this diagrammatically as a vertical line with time running from bottom to top. We mark $m$ points on the line, labeled from top to bottom by $\tau_1$ to $\tau_m$. Each vertex corresponds to a flip of the state of the detector, either from ground to excited state, or from excited to ground state. Thus, to each line segment between adjacent vertices corresponds a definite state of the detector. Moreover, these detector states alternate at each vertex. We depict the ground state by a straight line and the excited state by a dashed line. We can read off from expression (\ref{eq:udwintegrand}) and the formulas (\ref{eq:omegafact}), a factor for each vertex, depending on the detector transition at the vertex,
\begin{equation}
    g\to e:\quad -\im\lambda\chi(\tau) e^{\im\Omega\tau},\qquad
    e\to g:\quad -\im\lambda\chi(\tau) e^{-\im\Omega\tau} .
\end{equation}

\begin{figure}
    \centering
    \begin{tikzpicture}
    \begin{scope}
        \draw[-] (0,0.5) -- (0,2)node [midway,right] {g};
        \draw[dashed] (0,-1) -- (0,0.5)node [midway,right] {e};
        \draw [decorate, decoration={snake}] (0,0.5) -- (-2,0.5) node [left] {$\xi$};
        \draw[fill] (0,0.5) circle (0.1cm);
        \node at (-1,-1.5) {(a)};
    \end{scope}
    \begin{scope}[xshift=4cm]
        \draw[dashed] (0,0.5) -- (0,2)node [midway,right] {e};
        \draw[-] (0,-1) -- (0,0.5)node [midway,right] {g};
        \draw [decorate, decoration={snake}] (0,0.5) -- (-2,0.5) node [left] {$\xi$};
        \draw[fill] (0,0.5) circle (0.1cm);
        \node at (-1,-1.5) {(b)};
        \end{scope}
    \end{tikzpicture}
    \caption{Vertices corresponding to particle emission (a) and absorption (b).}
    \label{fig:vertices}
\end{figure}
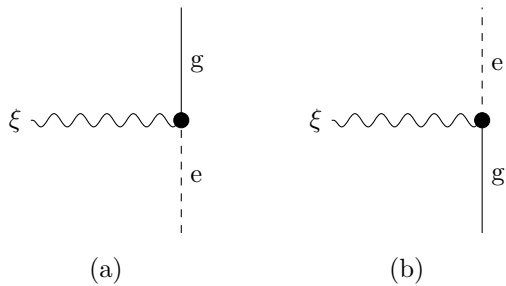

As usual, we depict the bosonic field $\phi$ by wavy lines. We read off from expression (\ref{eq:udwintegrand}) that a wavy line ends at each vertex, corresponding to the factor $\phi(x(\tau_k))$. The other end points of the wavy lines are the external particle lines. For an $n$ particle state $\Psi_{\xi_1,\dots,\xi_n}$ we have $n$ such external particle lines, labeled by the phase space elements $\xi_1,\dots,\xi_n$. This is illustrated in Figure~\ref{fig:vertices} with Feynman diagrams with a single vertex. In general, the wavy lines may connect:
\begin{enumerate}
    \item external particle lines with each other,
    \item a vertex with another vertex, or
    \item a vertex with an external particle line.
\end{enumerate}
In the first case we have a non-connected amplitude and obtain a factor $\{\xi_k,\xi_l^{\Lint}\}$ for the wavy line, as can be read off from formulas~(\ref{eq:freepairing}) and (\ref{eq:2ampl}), taking into account formula~(\ref{eq:decon}). In the second case we obtain a factor $\rho[\phi(x(\tau_k))\phi(x(\tau_l))](\Psi_{\Lvac})$ as can be read off from formula (\ref{eq:denorm}). This is just the Feynman propagator evaluated at the points $x(\tau_k)$ and $x(\tau_l)$. In the third case we obtain a factor of $\sqrt{2} \xi_k^{\Lint}(x(\tau_l))$ as can be read off from formulas (\ref{eq:nopair}) and (\ref{eq:partend}). Taking the product of the vertex factors and field-line factors we obtain an expression $A(\tau_1,\ldots,\tau_n)$ that depends on the times $\tau_1\ge\tau_2\ge\cdots\ge\tau_n$ assigned to the vertices. It remains to perform the interdependent time integrals, as can be read off from expression (\ref{eq:detobs}). This yields the amplitude associated to the Feynman diagram,
\begin{equation}
    \int_{-\infty}^{\infty}\xd \tau_1 \int_{-\infty}^{\tau_1}\xd \tau_2 \cdots \int_{-\infty}^{\tau_{n-1}}\xd \tau_n\, A(\tau_1,\ldots,\tau_n) .
\end{equation}

\subsection{Renormalization}
\label{sec:renorm}

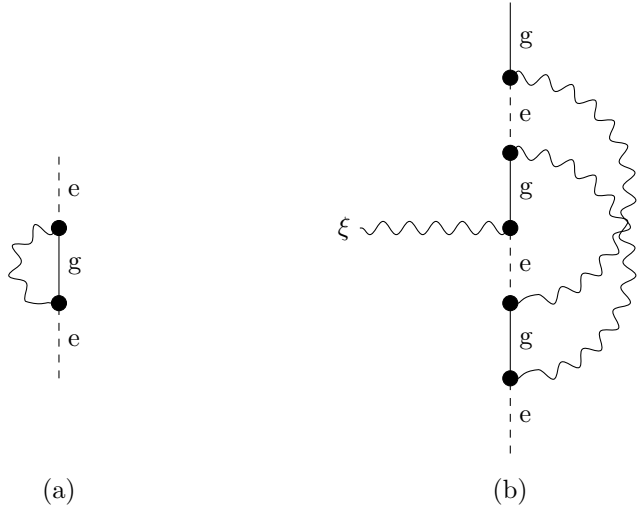
\begin{figure}
    \centering
    \begin{tikzpicture}
    \begin{scope}
        \draw[-] (0,0) -- (0,1)node [midway,right] {g};
        \draw[dashed] (0,-1) -- (0,0)node [midway,right] {e};
        \draw[dashed] (0,1) -- (0,2)node [midway,right] {e};
        \draw [decorate, decoration={snake}] (0,1) node [above left] {} arc (90:270:0.5)node [below left] {};
        \draw[fill] (0,0) circle (0.1cm);
        \draw[fill] (0,1) circle (0.1cm);
        \node at (0,-2.5) {(a)};
        \end{scope}
    \begin{scope}[xshift=6cm,yshift=-2cm]
        \draw[dashed] (0,0) -- (0,1)node [midway,right] {e};
        \draw[-] (0,1) -- (0,2)node [midway,right] {g};
        \draw[dashed] (0,2) -- (0,3)node [midway,right] {e};
        \draw[-] (0,3) -- (0,4)node [midway,right] {g};
        \draw[dashed] (0,4) -- (0,5)node [midway,right] {e};
        \draw[-] (0,5) -- (0,6)node [midway,right] {g};
        \draw [decorate, decoration={snake}] (0,3) -- (-2,3) node [left] {$\xi$};
        \draw[fill] (0,1) circle (0.1cm);
        \draw[fill] (0,2) circle (0.1cm);
        \draw[fill] (0,3) circle (0.1cm);
        \draw[fill] (0,4) circle (0.1cm);
        \draw[fill] (0,5) circle (0.1cm);
        \draw [decorate, decoration={snake}] (0,4) arc (90:-90:1.5);
        \draw [decorate, decoration={snake}] (0,5) arc (90:-90:1.5);
        \node at (0,-0.5) {(b)};
    \end{scope}
    \end{tikzpicture}
    \caption{Examples of diagrams renormalizing the propagator (a) and the vertex (b).}
    \label{fig:propvertexrenorm}
\end{figure}
    
Consider the Feynman diagrams (a) and (b) of Figure~\ref{fig:propvertexrenorm}. The first is a correction to the UDW propagator due to the interaction of the detector with the field vacuum. The second is a corresponding correction to the UDW vertex. There are more complicated corrections as well. In any case, already these simple corrections are divergent, when the involved vertices approach each other (in detector proper time). This is not surprising as we allow the detector to interact with the field vacuum repeatedly on arbitrarily short time scales. This was not the intended meaning when writing down the UDW Hamiltonians (\ref{eq:udwfreeham}) and (\ref{eq:udwintham}). These Hamiltonians were meant to describe the free evolution already with the field vacuum present, and the interaction with real (rather than virtual) particles only. That is, they should be regarded as describing the complete propagator and interaction, where the mentioned corrections are already taken into account and have been summed up. This is our renormalization prescription. UDW propagator and interaction already being complete translates in terms of diagrams simply into the absence of any diagram with wavy lines connecting vertices with each other. As can be seen from formula (\ref{eq:denorm}), this is precisely achieved if we impose \emph{normal ordered quantization} on the detector observable $O_{\psi_{\Lini}\to\psi_{\Lfin}}$ from the outset. We remark that the relation (\ref{eq:decon}) for the decomposition into connected amplitudes has an exact analog for normal ordered quantization,
\begin{equation}
    \rho[\no{O}](\Psi_{\xi_1,\dots,\xi_n})
    = \sum_{m=0}^{\lfloor n/2 \rfloor}  \sum_{\sigma\in S^n} \frac{1}{(2m)!\, (n-2m)!}\, \rho\left(\Psi_{\xi_{\sigma(1)},\dots,\xi_{\sigma(2m)}}\right)
    \rho_{\text{c}}[\no{O}]\left(\Psi_{\xi_{\sigma(2m+1)},\dots,\xi_{\sigma(n)}}\right) .
    \label{eq:ndecon}
\end{equation}
The evaluation of the normal ordered connected amplitude then proceeds directly with relation (\ref{eq:nopair}) while relation (\ref{eq:denorm}) is dropped. The Feynman rules are the same as before, while Feynman diagrams are restricted to not allow wavy lines connecting pairs of vertices. That is, the Feynman diagrams are tree-level only and may not contain any loops.


\section{Emission and absorption}
\label{sec:emabs}

In this paper we study the interaction of an inertial UDW detector with different particle states, with an emphasis on the novel notion of evanescent particle. Thus, the UDW detector remains at the origin, $x(\tau)=(\tau,\vec{0})$. What is more, it is switched on and off adiabatically via a switching function that we take to be a Gaussian,
\begin{equation}
    \chi(\tau)=\exp\left(-\pi\left(\frac{\tau}{T}\right)^2\right) .
\end{equation}
Note that this function satisfies the sufficient regularity conditions laid out in Appendix~\ref{sec:regularity}.
Here, $T$ is a timescale. It is chosen so that the integral over $\chi$ yields $T$. In other words, if the switching function was a characteristic function for a time interval, $T$ would be the duration of the interval. We can thus think of $T$ intuitively as the time duration for which the detector is switched on.

We start with the simplest process corresponding to emission or absorption of a single particle. We consider emission first. With the previously discussed Feynman rules, the amplitude for the transition of the UDW detector from the excited to the ground state with a single particle state on the hypercylinder is given by,
\begin{equation}
    \rho[O_{e\to g}](\Psi_{\xi})
    =-\im\sqrt{2} \lambda \int_{-\infty}^{\infty}\xd \tau\,
    \chi(\tau) e^{-\im \Omega\tau} \xi^{\Lint}(\tau,\vec{0}) .
    \label{eq:emampl}
\end{equation}
Similarly, for absorption, we obtain,
\begin{equation}
    \rho[O_{g\to e}](\Psi_{\xi})
    =-\im\sqrt{2} \lambda \int_{-\infty}^{\infty}\xd \tau\,
    \chi(\tau) e^{\im \Omega\tau} \xi^{\Lint}(\tau,\vec{0}) .
    \label{eq:absampl}
\end{equation}
Recall that $\xi^{\Lint}$ is a regular (i.e., well-defined in the interior of the hypercylinder), but in general complexified, solution of the Klein-Gordon equation determined by the particle state. Since $\xi^{\Lint}$ is evaluated at the origin in space, we can read off from the parametrization (\ref{eq:pparam}) and (\ref{eq:besseld}) that the coefficients $(\xi^{\Lint})^{\bullet}_{E,\ls,\ms}$ corresponding to nontrivial angular momentum $\ls\neq 0$ do not contribute. In other words, particles with nonvanishing angular momentum can neither be emitted nor absorbed by the detector. This is, of course, exactly what we expect due to the spherical symmetry of the system. In the following, we limit ourselves thus to external particles with vanishing angular momentum.
The remaining quantum numbers that we use are the energy and the binary in/out quantum number. For simplicity, we use the notation $\Psi^{\Lin}_E\defeq \Psi_{\Phi^{\Lin,E,0,0}}$ and $\Psi^{\Lout}_E\defeq \Psi_{\Phi^{\Lout,E,0,0}}$ for propagating 1-particle states and $\Psi^{\Lin}_E\defeq \Psi_{\tilde{\Phi}^{\Lin,E,0,0}}$ and $\Psi^{\Lout}_E\defeq \Psi_{\tilde{\Phi}^{\Lout,E,0,0}}$ for evanescent 1-particle states.

\subsection{Propagating particles}

We consider propagating particles first. With the particle states defined as in Section~\ref{sec:particles}, the amplitude for emission of an outgoing particle of energy $E$ is, (compare particularly equation (\ref{eq:propoutint})),
\begin{equation}
    \rho[O_{e\to g}](\Psi^{\Lout}_E)
    =-\frac{\im}{2\pi}\lambda\sqrt{p} \int_{-\infty}^{\infty}\xd \tau\,
    \chi(\tau) e^{\im (E-\Omega)\tau}  .
    \label{eq:amplemitpint}
\end{equation}
To evaluate this and similar integrals we recall the equality,
\begin{equation}
    \int_{-\infty}^{\infty}\xd \tau\,
    \chi(\tau) e^{\im s\tau}= T \exp\left(-\frac{s^2 T^2}{4\pi}\right) .
\end{equation}
As a function of $s$, this is a Gaussian peaked at $s=0$. The emission amplitude is thus,
\begin{equation}
    \rho[O_{e\to g}](\Psi^{\Lout}_E)
    =-\frac{\im}{2\pi}\lambda \sqrt{p}\, T \exp\left(-\frac{(E-\Omega)^2 T^2}{4\pi}\right) .
    \label{eq:amplemitp}
\end{equation}
The amplitude peaks when the particle energy $E$ coincides with the detector energy gap $\Omega$. On the grounds of conservation of energy we would expect the detector to emit a particle of energy exactly $E=\Omega$. That the amplitude and thus the probability of emission is non-zero, but exponentially suppressed for $E$ away from $\Omega$ is attributed to the unmodeled external influence that switches the detector on and off. Indeed, the peak of the amplitude increases in height and decreases in width when $T$ increases, i.e., when the switching is more adiabatic. In the limit of large $T$, we obtain a well-known delta function dependence,
\begin{equation}
    \lim_{T\to\infty} \rho[O_{e\to g}](\Psi^{\Lout}_E)
    =-\im \lambda \sqrt{p}\, \delta(E-\Omega) .
\end{equation}
Note that we have taken the particle to be outgoing, which is what we expect to be produced by the detector, recall Section~\ref{sec:inout}. If, in contrast, we take the particle to be incoming, we obtain the emission amplitude,
\begin{equation}
\rho[O_{e\to g}](\Psi^{\Lin}_E)
=-\frac{\im}{2\pi}\lambda\sqrt{p} \int_{-\infty}^{\infty}\xd \tau\,
\chi(\tau) e^{-\im (E+\Omega)\tau}
=-\frac{\im}{2\pi}\lambda \sqrt{p}\, T \exp\left(-\frac{(E+\Omega)^2 T^2}{4\pi}\right) .
\end{equation}
As expected, this is highly suppressed and vanishes in the limit $T\to\infty$. We consider this type of amplitude as unphysical, and an artifact of the simplicity of the model.

Considering single particle absorption instead of emission leads to precisely the same amplitudes, except for the interchange of incoming and outgoing quantum numbers. The discussion of particle energies is also analogous with absorption peaked where the incoming particle energy matches the detector energy gap. Absorption of outgoing particles is highly suppressed and vanishes in the limit of large $T$.

\subsection{Evanescent particles}

We proceed to consider the interaction of the UDW detector with evanescent particles. With the mathematical machinery in place, we can read off the amplitudes as in the propagating case. Thus, the amplitude for the emission of an outgoing evanescent particle is given by, (compare equation (\ref{eq:evanoutint})),
\begin{equation}
    \rho[O_{e\to g}](\Psi^{\Lout}_E)
    =-\frac{\im\, e^{\pi\im/4}}{2\pi}\lambda\sqrt{p} \int_{-\infty}^{\infty}\xd \tau\,
    \chi(\tau) e^{\im (E-\Omega)\tau}
    =-\frac{\im\, e^{\pi\im/4}}{2\pi}\lambda \sqrt{p}\, T \exp\left(-\frac{(E-\Omega)^2 T^2}{4\pi}\right) .
    \label{eq:amplemite}
\end{equation}
This looks strikingly similar to the amplitude for a propagating particle, compare equations (\ref{eq:amplemitpint}) and (\ref{eq:amplemitp}). Apparently, the difference is only a phase factor $e^{\pi\im/4}$. However, there is another important difference. In the evanescent case, $p$ is not the particle momentum, but $p=\sqrt{m^2-E^2}$. (In both cases $p=\sqrt{|E^2-m^2|}$.)

The first and most important result here is that, yes, evanescent particles are emitted by the detector and carry away energy. Moreover, they do so in a manner very similar to propagating particles. The amplitude peaks when the particle energy coincides with the detector energy gap. For this to occur the detector energy gap has to be smaller than the mass of the field. Also, as in the propagating case, in the infinite time limit $T\to\infty$ the amplitude becomes a delta-function.
If we replace the outgoing particle with an incoming one, the amplitude becomes,
\begin{equation}
    \rho[O_{e\to g}](\Psi^{\Lin}_E)
    =-\frac{\im\, e^{\pi\im/4}}{2\pi}\lambda \sqrt{p}\, T \exp\left(-\frac{(E+\Omega)^2 T^2}{4\pi}\right) .
\end{equation}
This is again completely analogous to the propagating case with an amplitude that is highly suppressed. What is more, the amplitudes for absorption are exactly the same as those for emission, except for the interchange of "incoming" and "outgoing" quantum numbers.

As a second significant result we thus find that the evanescent particles can indeed be meaningfully characterized as "incoming" and "outgoing" in the sense of transporting energy into or out of the hypercylinder, just like their propagating counterparts. This might seem surprising from the point of view of an intuition trained on situations in classical physics where evanescent waves are associated with a lack of transport of energy. However, as seen in Section~\ref{sec:inout}, this behavior has an exact classical counterpart. 

This leads us to the third result we would like to stress. That is, the modes labeled as "incoming" or "outgoing" in the classical theory (Section~\ref{sec:inout}) are precisely mapped to the quantum particle states with the corresponding property, not only in the propagating, but also in the evanescent sector. And this is the case in spite of significant differences in the quantization scheme between the two sectors. In other words, the novel twisted Kähler quantization scheme is "correct" not only for the energy quantum number (which is easy to achieve due to time-translation symmetry), but also for the much more intricate "incoming"/"outgoing" quantum number. This appears to be a nontrivial result.


\section{Emission spectrum}
\label{sec:espectrum}

In the present section we consider the emission spectrum of the UDW detector. That is, we quantify the probability density of particle emission as a function of the particle energy. The precise assumptions are the following:
\begin{itemize}
    \item The detector is at early times in the excited state and at late times in the ground state.
    \item There is exactly one particle in the boundary Hilbert space. The particle is outgoing and has exact energy and angular momentum quantum numbers.
\end{itemize}
As we saw previously, the amplitude vanishes if the angular momentum of the particle is nonvanishing, so we can restrict to consider the energy as the only explicit variable. With the completeness relation (\ref{eq:hs1compl}) restricted to the outgoing sector, we obtain the probability per unit energy,
\begin{equation}
  P(E)=\frac{\left|\rho[O_{e\to g}](\Psi^{\Lout}_E)\right|^2}{\int_0^\infty \xd E'\,\left|\rho[O_{e\to g}](\Psi^{\Lout}_{E'})\right|^2} .
  \label{eq:specexpem}
\end{equation}
With expressions (\ref{eq:amplemitp}) and (\ref{eq:amplemite}) this is
\begin{equation}
    P(E)=\frac{p(E) \exp\left(-\frac{(E-\Omega)^2 T^2}{2\pi}\right)}{\int_0^\infty \xd E'\, p(E') \exp\left(-\frac{(E'-\Omega)^2 T^2}{2\pi}\right)} .
    \label{eq:specem}
\end{equation}
We emphasize that within the assumptions made, this expression is exact. Due to the fixed particle number and the renormalization, there are no other diagrams that contribute.
Apart from the peak of the spectrum at the detector energy gap $\Omega$ that we have already discussed in the context of the amplitude, there is another interesting feature of the spectrum. Namely, the momentum factor $p$ causes a suppression when the particle energy $E$ is close to the mass $m$.

\begin{figure}
    \includegraphics[width=0.50\textwidth]{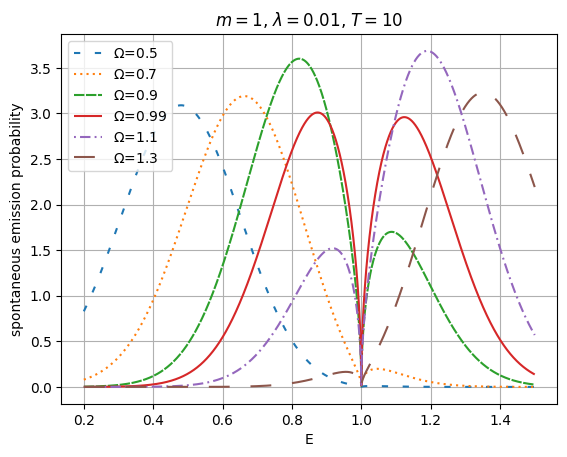}%
    \includegraphics[width=0.50\textwidth]{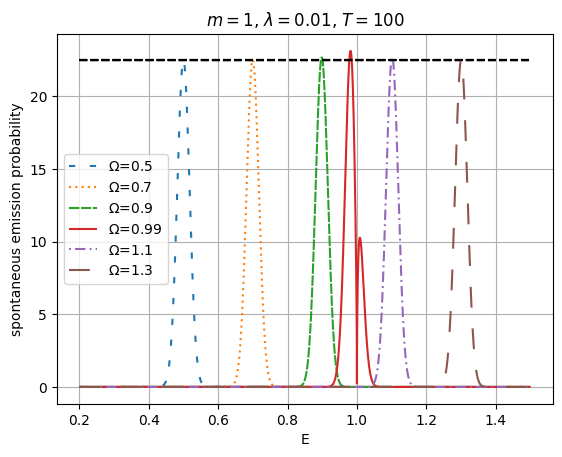}
    \caption{Emission spectrum, namely probability per unit energy, for different values of the detector energy gap $\Omega$ expressed in units of the mass of the field. In the left-hand (right-hand) plot the time $T$ takes the value $10$ ($100$). The coupling constant $\lambda$ has been set equal to $0.01$.}
    \label{fig:es}
\end{figure}

Figure~\ref{fig:es} shows plots of the emission spectrum for different values of the detector energy gap $\Omega$ and time $T$. The peak of the spectrum at coincidence of particle energy with the detector gap is clearly visible. Moreover, we can appreciate that the peak becomes narrower when the detector time $T$ increases, compare the left-hand plot ($T=10$) to the right-hand one ($T=100$). In the adiabatic limit $T\to\infty$, the probability density is a delta function,
\begin{equation}
    \lim_{T\to\infty} P(E)=\delta(E-\Omega) .
\end{equation}
We can also appreciate in the plots the mentioned suppression of emission when the particle energy is close to the mass of the field. This is more clearly visible at relatively shorter detector times due to the larger width of the spectrum (here at $T=10$ compared to $T=100$).
Another feature we observe is that for both values of $T$ the maximum of the probability density is independent of $\Omega$, both in the evanescent sector ($\Omega <m$) and the propagating one ($\Omega >m$). To see this, we approximate (\ref{eq:specem}) for relatively large $T$ by the Laplace method, yielding,
\begin{equation}
    P(E) \approx \frac{T}{\sqrt{2} \pi}\frac{p(E) \exp\left(-\frac{(E-\Omega)^2 T^2}{2\pi}\right)}{p(\Omega)} .
\end{equation}
This shows that the maximum of the probability, achieved when the energy of the outgoing particle equals the detector energy gap, is given by $\frac{T}{\sqrt{2} \pi}$. This is indicated by the horizontal dashed line in the right-hand plot of Figure~\ref{fig:es}.


\section{Spontaneous emission probability}
\label{sec:eprob}

We proceed to consider the total emission probability. In contrast to the considerations of the previous section this means that we allow for the possibility that no particle is emitted. The precise assumptions are the following:
\begin{itemize}
    \item At early times the detector is in the excited state. The late-time state of the detector is unknown.
    \item There are only outgoing particles in the boundary Hilbert space.
    \item We exclude as spurious contributions that involve an outgoing particle being absorbed by the detector. (Compare the discussion in Section~\ref{sec:emabs}.)
\end{itemize}
There are precisely two processes allowed by the assumptions. Firstly, this is the transition of the detector from the excited to the ground state while a single outgoing particle is emitted, as in Section~\ref{sec:espectrum}. Secondly, this is the detector remaining in the excited state while the field is in the vacuum. The total spontaneous emission probability is thus given as follows:
\begin{equation}
    P=\frac{\int_0^\infty \xd E\,\left|\rho[O_{e\to g}](\Psi^{\Lout}_E)\right|^2}{\left|\rho[O_{e\to e}](\Psi_{\Lvac})\right|^2+\int_0^\infty \xd E\,\left|\rho[O_{e\to g}](\Psi^{\Lout}_E)\right|^2} .
    \label{eq:probexpse}
\end{equation}
Inserting expressions (\ref{eq:amplemitp}) and (\ref{eq:amplemite}), we get
\begin{equation}
    P=\frac{\frac{\lambda^2 T^2}{4\pi^2}\int_0^\infty \xd E\, p \exp\left(-\frac{(E-\Omega)^2 T^2}{2\pi}\right)}{1+\frac{\lambda^2 T^2}{4\pi^2}\int_0^\infty \xd E\, p \exp\left(-\frac{(E-\Omega)^2 T^2}{2\pi}\right)}
    =\frac{1}{1+\frac{4\pi^2}{\lambda^2 T^2\int_0^\infty \xd E\, p \exp\left(-\frac{(E-\Omega)^2 T^2}{2\pi}\right)}} .
    \label{eq:probse}
\end{equation}
For large $T$ we can approximate the integral by the Laplace method, yielding,
\begin{equation}
    P\approx \frac{1}{1+\frac{2\sqrt{2}\pi}{\lambda^2 T p(\Omega)}} .
\end{equation}

\begin{figure}
    \includegraphics[width=0.48\textwidth]{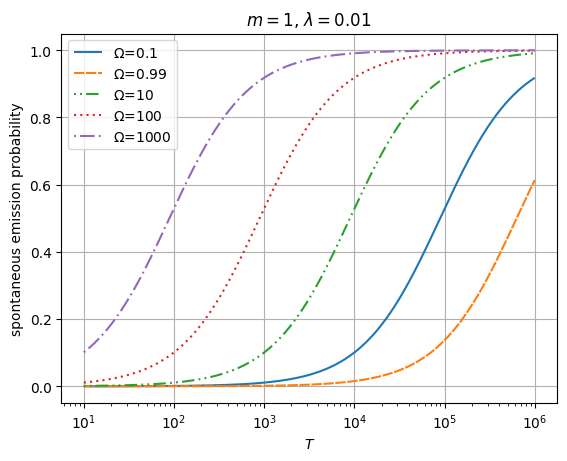}%
    \includegraphics[width=0.50\textwidth]{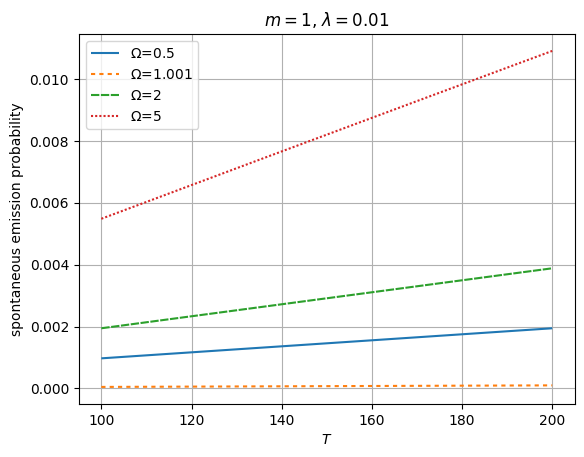}
    \caption{Spontaneous emission probability as a function of the time parameter $T$, for different values of the detector energy gap $\Omega$ expressed in unit mass of the field, at $\lambda=0.01$. On the right-hand side, the focus is on relatively small values of $T$, with an appreciably linear dependence of the probability on $T$.}
    \label{fig:sep-T}
\end{figure}

A key characteristic quantity for spontaneous emission is the \emph{emission rate} $R$. This is the probability of emission of a detector evolving from an initial excited state, per unit time. In the graphs of $P$ as a function of $T$ this corresponds to the slope of the curve in the regime of low probability, where it approximates a straight line. We can easily extract this from formula (\ref{eq:probse}) by pretending $\lambda$ to be small, then dividing by $T$, and then taking the adiabatic limit $T\to\infty$. This yields,
\begin{equation}
    R=\frac{\lambda^2 p(\Omega)}{2\sqrt{2}\pi} .
\end{equation}
In Figure~\ref{fig:sep-T}, the probability of spontaneous emission is shown as a function of $T$. In the regime where the probability is low (right-hand plot), the linear dependence with approximately $P= R T$ can be clearly appreciated. In fact, in this regime the curves for $P=R T$ are not visually distinguishable from the true curves given by equation \eqref{eq:probse}.

\begin{figure}
    \begin{center}
    \includegraphics[width=0.6\textwidth]{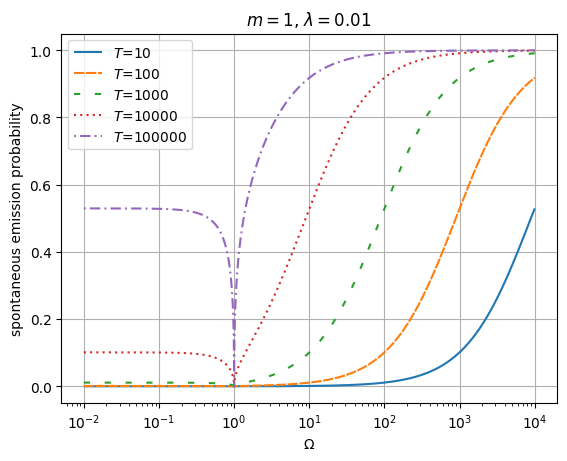}
    \caption{Spontaneous emission probability as a function of the detector energy gap $\Omega$ expressed units of mass of the field for different values of $T$, at $\lambda=0.01$.}
    \label{fig:sep-Omega}
    \end{center}
\end{figure}

We note that the emission probability is \emph{not} given by an exponential decay law, determined by the rate $R$, which would give $P=1-\exp(-RT)$. The actual increase of the probability of emission with increased detector interaction time until saturation at unit probability can be observed in the left-hand plot shown in Figure~\ref{fig:sep-T}. Comparing different detector energy gaps shows that higher-energy gaps lead to a higher emission probability in the propagating sector. In the evanescent sector it is lower energies that lead to higher probability. When the energy gap equals the mass of the field, particle emission becomes suppressed and its probability vanishes in the adiabatic limit $T\to\infty$. This is also clearly visible in the plot of Figure~\ref{fig:sep-Omega}, showing the probability as a function of the energy gap for different fixed values of the time $T$.

\begin{figure}
    \includegraphics[width=0.48\textwidth]{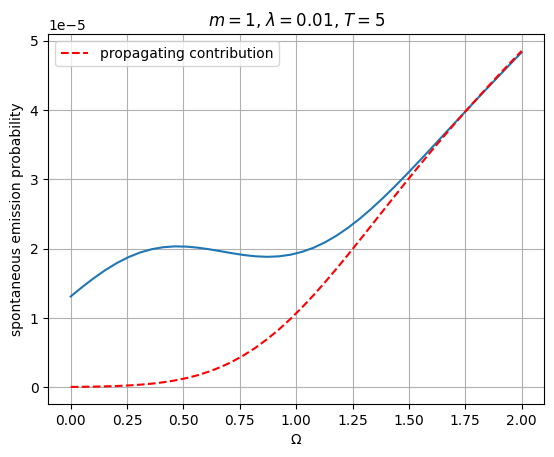}%
    \includegraphics[width=0.52\textwidth]{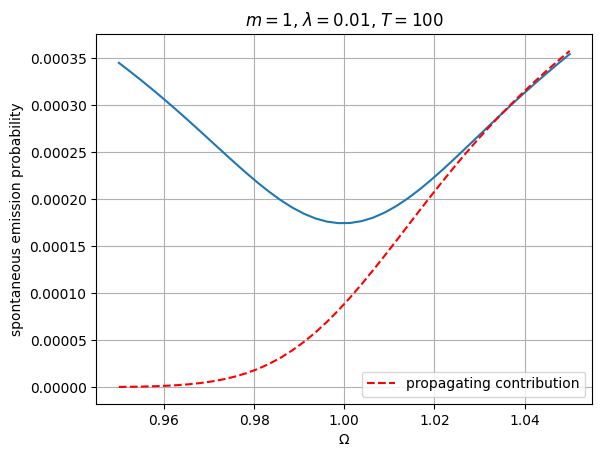}
    \caption{Spontaneous emission probability as a function of the detector energy gap $\Omega$, at $\lambda=0.01$. In addition to the emission probability for the radial picture (solid line), the emission probability for the temporal picture (dashed line) is also indicated. The characteristic time $T$ is $5$ (left-hand plot) and is $100$ (right-hand plot).}
    \label{fig:sep-radtemp}
\end{figure}

It is instructive to compare the emission spectrum for the present radial picture with that which would be obtained for the temporal picture. As explained previously, the probabilities for the latter are precisely obtained by removing the evanescent sector. That is, the integrals over the energy in numerator and denominator of expression (\ref{eq:probexpse}) restrict in this case to the range $E>m$. In Figure~\ref{fig:sep-radtemp} the emission probabilities for both pictures are compared as a function of the detector gap energy $\Omega$ near the field mass, i.e., near $\Omega=m$. At $\Omega < m$ it is not surprising that the results are different, because in the radial picture the detector can decay by emitting evanescent particles with energy less than $m$, while in the temporal picture it cannot. However, even at $\Omega>m$, the difference is notable if $\Omega-m$ is small. This "spillover" effect of the evanescent sector into the propagating sector is linked to the nonadiabaticity of the detector switching. That is, as $T$ is taken to be larger, $\Omega$ has to be increasingly closer to $m$ for the effect to be noticeable.


\section{Absorption probability}
\label{sec:aprob}

In this section we consider the probability of absorption of a single particle by the UDW detector as a function of the particle energy.

\subsection{Assumptions and probability formula}

Our assumptions this time are the following:
\begin{itemize}
    \item At early times the detector is in the ground state. The late-time state of the detector is unknown.
    \item There is precisely one incoming particle in the boundary Hilbert space. The outgoing sector of the boundary Hilbert space is unknown.
    \item We disregard as spurious contributions that involve an outgoing particle being absorbed by the detector.
\end{itemize}

\begin{figure}
    \begin{center}
    \begin{tikzpicture}
    \begin{scope}
    \draw[dashed] (0,0.5) -- (0,2)node [midway,right] {e};
    \draw[-] (0,-1) -- (0,0.5)node [midway,right] {g};
    \draw [decorate, decoration={snake}] (0,0.5) -- (-2,0.5) node [left] {in};
    \draw[fill] (0,0.5) circle (0.1cm);
    \node at (-1,-1.5) {(a)};
    \end{scope}
    \begin{scope}[xshift=5cm]
    \draw[dashed] (0,0) -- (0,1)node [midway,right] {e};
    \draw[-] (0,-1) -- (0,0)node [midway,right] {g};
    \draw [decorate, decoration={snake}] (0,0) -- (-2,0)node [left] {in};
    \draw[fill] (0,0) circle (0.1cm);
    \draw[-] (0,1) -- (0,2)node [midway,right] {g};
    \draw [decorate, decoration={snake}] (0,1) -- (-2,1)node [left] {out};
    \draw[fill] (0,1) circle (0.1cm);
    \node at (-1,-1.5) {(b)};
    \end{scope}
    \begin{scope}[xshift=10cm]
    \draw[-] (0,-1) -- (0,2)node [midway,right] {g};
    \draw [decorate, decoration={snake}] (-2,-0.5) node [left] {in} arc (-90:91:1.1)node [left] {out};
    \node at (-1,-1.5) {(c)};
    \end{scope}
    \end{tikzpicture}
    \caption{Feynman diagrams for processes relevant to absorption.}
    \label{fig:3Fd}
    \end{center}
\end{figure}
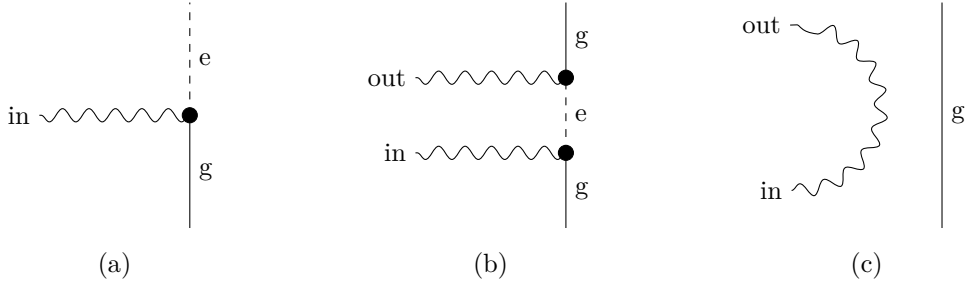

There are three processes that contribute:
\begin{enumerate}
\item[(a)] The incoming particle is absorbed by the detector and the detector ends up in the excited state.
\item[(b)] The incoming particle is absorbed by the detector, but subsequently the detector emits an outgoing particle. The detector ends up in the ground state.
\item[(c)] The incoming particle does not interact with the detector, but leaves as an outgoing particle. The detector remains in the ground state.
\end{enumerate}
The corresponding Feynman diagrams are depicted in Figure~\ref{fig:3Fd}. The probability of absorption for a particle characterized by a phase space element $\xi\in L^{\Lp,\Lin}\oplus L^{\Le,\alpha,\Lin}$ is
\begin{equation}
    P=\frac{|\rho[O_{g\to e}](\Psi_{\xi})|^2}{|\rho[O_{g\to e}](\Psi_{\xi})|^2
     + \int\xd E\, |\rho[O_{g\to g}](\Psi_{\xi}\tens\Psi^{\Lout}_{E})|^2} .
    \label{eq:absprob}
\end{equation}
The first term in the denominator corresponds to process (a), while the second term comprises both processes (b) and (c).
From equation (\ref{eq:ndecon}) we obtain the decomposition of the amplitude in the second term of the denominator into connected and disconnected parts, corresponding to processes (b) and (c) respectively,\footnote{Strictly speaking, the connected amplitude also includes a version of process (b), where the roles of the incoming and outgoing particles are interchanged. However, this terms is very highly suppressed, considered as spurious in agreement with the stated assumptions, and ignored in the following.}
\begin{equation}
    \rho[O_{g\to g}](\Psi_{\xi}\tens\Psi^{\Lout}_{E})
    = \rho_{\text{c}}[O_{g\to g}](\Psi_{\xi}\tens\Psi^{\Lout}_{E}) 
     + \rho(\Psi_{\xi}\tens\Psi^{\Lout}_{E}) .
     \label{eq:bcsep}
\end{equation}
It is useful to separate the different terms arising when taking the modulus square and integrating over the right-hand side of equation (\ref{eq:bcsep}). We denote these terms by [b], [c], and [m]. Here, [b] and [c] arise from the modulus square of the amplitudes for (b) and (c) respectively, while [m] denotes the mixed term involving both amplitudes (b) and (c). We also call [a] the term arising as the modulus square of the amplitude for (a).

\subsection{Relations and simplifications}

Before proceeding to evaluate the different terms, we look for relations and simplifications. The mixed term [m] is,
\begin{equation}
    \int\xd E\, \rho_{\text{c}}[O_{g\to g}](\Psi_{\xi}\tens\Psi^{\Lout}_{E})
    \overline{\rho(\Psi_{\xi}\tens\Psi^{\Lout}_{E})} + \text{c.c.}
\end{equation}
The factor consisting of the free amplitude can be evaluated with (\ref{eq:2ampl}) as follows:
\begin{equation}
    \overline{\rho(\Psi_{\xi}\tens\Psi^{\Lout}_{E})}=\overline{\{\Phi^{\Lout,E,0,0},u(\xi)\}}=\{u(\xi),\Phi^{\Lout,E,0,0}\} .
\end{equation}
By linearity of the amplitude we can take this as a factor multiplying the state, allowing us to resolve the integral as a completeness relation,
\begin{equation}
    \int\xd E\, \{u(\xi),\Phi^{\Lout,E,0,0}\} \Psi^{\Lout}_{E} 
    = \Psi_{u(\xi)} .
\end{equation}
We have used here that $\xi$ encodes an incoming particle with vanishing angular momentum. Thus, the mixed term [m] is
\begin{equation}
    \rho_{\text{c}}[O_{g\to g}](\Psi_{\xi}\tens\Psi_{u(\xi)})
     + \text{c.c.}
\end{equation}
There is a way to read this off, combining the amplitude of process (b) with the conjugate one for process (c) from the Feynman diagrams. Concretely, relative conjugation and the completeness relation imply that we can glue together the two external legs labeled "out" in Figure~\ref{fig:3Fd}.(b) and (c) which both carry the particle state $\Psi^{\Lout}_{E'}$ that is integrated out. The detector line in Figure~\ref{fig:3Fd}.(c) is inert and may simply be ignored. The result is a diagram identical to that of Figure~\ref{fig:3Fd}.(b), but with the leg labeled "out" carrying now the "dualized" version (due to complex conjugation) of the state originally labeling the "in" leg in Figure~\ref{fig:3Fd}.(c).

We may notice that the Feynman diagram of process (b) looks like the diagram of process (a) glued together with a mirror image copy of itself on top (Figure~\ref{fig:3Fd}). It turns out that we can transform also this statement into a statement about the corresponding amplitudes. The term [a] corresponding to the modulus square of the amplitude for process (a) takes the form (recall equation (\ref{eq:absampl})),
\begin{multline}
    \rho[O_{g\to e}](\Psi_{\xi}) \overline{\rho[O_{g\to e}](\Psi_{\xi})} \\
    = \left(-\im\sqrt{2} \lambda \int_{-\infty}^{\infty}\xd \tau\,
    \chi(\tau) e^{\im \Omega\tau} \xi^{\Lint}(\tau,\vec{0})\right)
    \left(+\im\sqrt{2} \lambda \int_{-\infty}^{\infty}\xd \tau\,
    \chi(\tau) e^{-\im \Omega\tau} \overline{\xi^{\Lint}}(\tau,\vec{0})\right)
    \label{eq:expra} .
\end{multline}
Observe that, except for the overall sign and the replacement of $\xi^{\Lint}$ by its conjugate, the second factor is identical to the amplitude for the emission process, expression~(\ref{eq:emampl}). On the other hand, the term [m] takes the form,
\begin{multline}
    \rho_{\text{c}}[O_{g\to g}](\Psi_{\xi}\tens\Psi_{u(\xi)}) + 
    \overline{\rho_{\text{c}}[O_{g\to g}](\Psi_{\xi}\tens\Psi_{u(\xi)})} \\
    =-2 \lambda^2 \int_{-\infty}^{\infty}\xd \tau\,
     \int_{-\infty}^{\tau}\xd \tau'\, \chi(\tau)\chi(\tau') e^{\im \Omega(\tau'-\tau)}\xi^{\Lint}(\tau',\vec{0})(u(\xi))^{\Lint}(\tau,\vec{0}) \\
    -2 \lambda^2 \int_{-\infty}^{\infty}\xd \tau\,
     \int_{-\infty}^{\tau}\xd \tau'\, \chi(\tau)\chi(\tau') e^{\im \Omega(\tau-\tau')}\overline{\xi^{\Lint}}(\tau',\vec{0})\overline{(u(\xi))^{\Lint}}(\tau,\vec{0}) .
      \label{eq:mterm}
\end{multline}
Given that $\xi$ is assumed to have vanishing angular momentum, one can show, using the relations of Appendix~\ref{sec:extraquant},
\begin{equation}
    (u(\xi))^{\Lint}=c\, \overline{\xi^{\Lint}} .
\end{equation}
Here $c=1$ in the propagating case and $c=\im$ in the evanescent case. We may thus rewrite the term [m] as
\begin{multline}
-2 c \lambda^2 \int_{-\infty}^{\infty}\xd \tau\,
     \int_{-\infty}^{\tau}\xd \tau'\, \chi(\tau)\chi(\tau') e^{\im \Omega(\tau'-\tau)}\xi^{\Lint}(\tau',\vec{0})\overline{\xi^{\Lint}}(\tau,\vec{0}) \\
-2 \overline{c} \lambda^2 \int_{-\infty}^{\infty}\xd \tau\,
     \int_{\tau}^{\infty}\xd \tau'\, \chi(\tau)\chi(\tau') e^{\im \Omega(\tau'-\tau)}\xi^{\Lint}(\tau',\vec{0})\overline{\xi^{\Lint}}(\tau,\vec{0})
     \label{eq:[m]} .
\end{multline}
In the second term we have also interchanged $\tau$ and $\tau'$ and adapted the integral accordingly. This makes it manifest that in the propagating case $c=1$ the integration ranges of the two terms precisely combine to give exactly the expression [a] of equation (\ref{eq:expra}), except for an opposite overall sign. That is, in the propagating case, the terms [a] and [m] precisely cancel out. In the evanescent case, this is not so.

We want to consider the absorption probability (\ref{eq:absprob}) as a function of the energy of the incoming particle. The most straightforward implementation of this would be to take an energy eigenstate, as in previous sections. However, the term [c] yields a singular contribution in this case. Indeed, [c] is
\begin{multline}
    \int\xd E\, |\rho(\Psi_\xi\tens\Psi^{\Lout}_{E})|^2
    = \int\xd E\, |\{\Phi^{\Lout,E,0,0},u(\xi)\}|^2 \\
    = \int\xd E\, \{u(\xi),\Phi^{\Lout,E,0,0}\}
    \{\Phi^{\Lout,E,0,0},u(\xi)\} 
    = \{u(\xi),u(\xi)\}
    = \{\xi,\xi\} =\|\Psi_{\xi}\|^2.
    \label{eq:termc}
\end{multline}
We have used here again that $\xi$ encodes an incoming particle with vanishing angular momentum. We see that this contribution is the square of the norm of $\xi$ in $L^{\Lp}\oplus L^{\Le,\alpha}$. But for the particle states with sharp energy defined in Sections~\ref{sec:tparticles} and \ref{sec:particles} this diverges as they are $\delta$-function normalized.
Physically this means that a particle with an exact sharp energy cannot be absorbed by the detector, but will always "miss" the detector.\footnote{To be clear, the expected divergence of the term [c] for a state of infinite norm alone is not enough to merit this physical interpretation. Rather, this interpretation arises by taking this together with the fact that the term [a] is only finite for the sharp energy state of infinite norm.}

To address this situation, we introduce states with a Gaussian energy spread $\Delta$ around a central energy value $E$. Also, we avoid superposing propagating and evanescent particles. That is, we cut off the energy spread at $E=m$ so that any particle we consider is either purely propagating or purely evanescent.
\begin{align}
    \Psi^{\Lin}_{E,\Delta}\defeq c_{E,\Delta}\int_m^\infty \xd E' \exp\left(-\pi\left(\frac{E-E'}{\Delta}\right)^2\right)\Psi^{\Lin}_{E'} & \qquad\text{if}\quad E>m , \\
    \Psi^{\Lin}_{E,\Delta}\defeq c_{E,\Delta}\int_0^m \xd E' \exp\left(-\pi\left(\frac{E-E'}{\Delta}\right)^2\right)\Psi^{\Lin}_{E'} & \qquad\text{if}\quad E<m .
\end{align}
The constants $c_{E,\Delta}$ may be chosen so that the states $\Psi^{\Lin}_{E,\Delta}$ are normalized. However, they are irrelevant for the probability (\ref{eq:absprob}) as they cancel out.

\subsection{Evaluation}

We are now ready to evaluate the absorption probability (\ref{eq:absprob}) for an incoming particle state $\Psi^{\Lin}_{E,\Delta}$. We first suppose that the particle is propagating and consider the relevant terms in turn.
The term [a] reads,
\begin{multline}
   \rho[O_{g\to e}](\Psi^{\Lin}_{E,\Delta}) \overline{\rho[O_{g\to e}](\Psi^{\Lin}_{E,\Delta})} \\
    = |c_{E,\Delta}|^2 \int_m^\infty \xd E' \, \xd E'' e^{-\pi \frac{(E-E')^2+(E-E'')^2}{\Delta^2}}
    \rho[O_{g\to e}](\Psi^{\Lin}_{E'}) \overline{\rho[O_{g\to e}](\Psi^{\Lin}_{E''})} .
    \label{eq:expra-p}
\end{multline}
The absorption amplitude $\rho[O_{g\to e}](\Psi^{\Lin}_{E})$ is easily seen to be the same as the corresponding emission amplitude (\ref{eq:amplemitp}),
\begin{equation}
  \rho[O_{g\to e}](\Psi^{\Lin}_{E}) = -  \frac{\im}{2 \pi} \lambda\sqrt{p}\, T  \exp\left(- \frac{(E- \Omega)^2 T^2}{4 \pi} \right).
  \label{eq:expra-p2}
\end{equation}
The integrals over $E'$ and $E''$ can be evaluated using the Laplace method, as reported in Appendix \ref{sec:app-ap-a}, valid in the regime,
\begin{equation}
\frac{4 \pi^2+T^2\Delta^2}{4 \pi \Delta^2} \gg1 .
\end{equation}
Note that we are in this regime if either $\Delta\ll 1$ or $T\gg 1$. We obtain,
\begin{multline}
   \rho[O_{g\to e}](\Psi^{\Lin}_{E,\Delta}) \overline{\rho[O_{g\to e}](\Psi^{\Lin}_{E,\Delta})} \\
   \simeq   \frac{\lambda^2 T^2 \Delta^2}{4\pi^2 + T^2 \Delta^2}  |c_{E,\Delta}|^2   \exp \left(  - \frac{ 2\pi T^2 }{4 \pi^2 +T^2 \Delta^2} (E-\Omega)^2\right) p\left( \frac{4 \pi^2 E + T^2 \Delta^2 \Omega}{4 \pi^2+T^2\Delta^2 }\right) .
\end{multline}

Recall that the term [b] is given here by
\begin{equation}
\int\xd E'\, |\rho_{\text{c}}[O_{g\to g}](\Psi^{\Lin}_{E,\Delta}\tens\Psi^{\Lout}_{E'})|^2 .
\label{eq:bterm}
\end{equation}
In this expression the amplitude of the process (b) is
\begin{equation}
\rho_{\text{c}}[O_{g\to g}](\Psi^{\Lin}_{E,\Delta}\tens\Psi^{\Lout}_{E'})
=c_{E,\Delta} \int_m^{\infty} \xd E'' e^{ -\pi  \left(\frac{E-E''}{\Delta} \right)^2}  \rho_{\text{c}}[O_{g\to g}](\Psi^{\Lin}_{E''}\tens\Psi^{\Lout}_{E'}) ,
\label{eq:a-p-[b]}
\end{equation}
where
\begin{align}
&\rho_{\text{c}}[O_{g\to g}](\Psi^{\Lin}_{E''}\tens\Psi^{\Lout}_{E'}) \nonumber\\
&=- 2\lambda^2
\int^{\infty}_{-\infty} \xd \tau  \, \chi(\tau) e^{-\im \Omega \tau}  (\Phi^{\Lout,E',0,0})^{\Lint}(\tau,0) 
  \int^{\tau}_{-\infty} \xd {\tau'} \, \chi(\tau') e^{\im \Omega \tau'}  (\Phi^{\Lin,E'',0,0})^{\Lint}(\tau',0).
\end{align}
In Appendix~\ref{sec:app-ap-b} the integrals are evaluated yielding the result
\begin{multline}
\rho_{\text{c}}[O_{g\to g}](\Psi^{\Lin}_{E,\Delta}\tens\Psi^{\Lout}_{E'})
  \simeq -  \frac{\lambda^2  \Delta T^2}{4  \pi \sqrt{4 \pi^2 + T^2 \Delta^2}}  c_{E,\Delta}e^{-\frac{\pi T^2 }{4 \pi^2 + T^2 \Delta^2} (E-\Omega)^2}   e^{-\frac{(E'-\Omega)^2}{4 \pi} T^2}
\\
  \sqrt{p(E')p\left(  \frac{4 \pi^2E + T^2 \Delta^2 \Omega}{4 \pi^2 + T^2 \Delta^2}\right)}  \left( \mathrm{erf} \left( \im \frac{T}{2 \sqrt{2 \pi}}\left[ \frac{4\pi^2(E-\Omega)}{4\pi^2+T^2 \Delta^2}  + E'- \Omega\right]\right)+1\right).
 \end{multline}
The integral over $E'$ of the modulus square of the above amplitude, corresponding to the term [b] of \eqref{eq:bterm} is evaluated in the regime $T \gg 1$ with the Laplace method, yielding
\begin{multline}
\int_m^{\infty} \xd E' |\rho_{\text{c}}[O_{g\to g}](\Psi^{\Lin}_{E,\Delta}\tens\Psi^{\Lout}_{E'})|^2 
  \simeq   \frac{\lambda^4 \Delta^2 T^3 \sqrt{2} \pi}{16 \pi^2 (4 \pi^2 + T^2 \Delta^2)}  |c_{E,\Delta}|^2 e^{-\frac{2\pi T^2 }{4 \pi^2 + T^2 \Delta^2} (E-\Omega)^2} \\
  p(\Omega) p\left(  \frac{4 \pi^2E + T^2 \Delta^2 \Omega}{4 \pi^2 + T^2 \Delta^2}\right) 
 \left( 1 -
\mathrm{erf}^2 \left( \im \frac{T}{2 \sqrt{2 \pi}}\left[ \frac{4\pi^2(E-\Omega)}{4\pi^2+T^2 \Delta^2}  \right]\right)\right).
 \label{eq:ms-b}
\end{multline}

The term [c] is just the norm square of the state $\Psi^{\Lin}_{E,\Delta}$ due to \eqref{eq:termc}, which we take to be unity,
\begin{equation}
    \int\xd E'\, |\rho(\Psi^{\Lin}_{E,\Delta}\tens\Psi^{\Lout}_{E'})|^2
    = \|\Psi^{\Lin}_{E,\Delta}\|^2=1 .
\end{equation}
Conversely, we calculate the normalization constant $c_{E,\Delta}$,
\begin{multline}
    \frac{1}{|c_{E,\Delta}|^{2}} = \int_m^\infty \xd E' \, \xd E'' e^{-\pi \frac{(E-E')^2+(E-E'')^2}{\Delta^2}} \{\Phi^{\Lin,E',0,0},\Phi^{\Lin,E'',0,0}\} \\
    = \int_m^\infty \xd E' e^{-\pi \frac{2 (E-E')^2}{\Delta^2}}
    \simeq \frac{\Delta}{\sqrt{2}}.
\end{multline}
In the last step, we have approximated the integral by removing the energy cutoff. This is valid if $\Delta\ll E-m$.

Combining all the ingredients, the absorption probability of an incoming propagating particle takes the form
\begin{equation}
P \simeq \frac{1   
}{
 \frac{\lambda^2 T \sqrt{2} \pi}{16 \pi^2}   p(\Omega) 
  \left( 1 -
\mathrm{erf}^2 \left( \im \frac{T}{2 \sqrt{2 \pi}}\left[ \frac{4\pi^2(E-\Omega)}{4\pi^2+T^2 \Delta^2}  \right]\right)\right)
+
\frac{4\pi^2 + T^2 \Delta^2}{\lambda^2 T^2 \Delta}
\frac{ e^{\frac{2\pi T^2 (E-\Omega)^2 }{4 \pi^2 + T^2 \Delta^2}}}{\sqrt{2} p\left( \frac{4 \pi^2 E + T^2 \Delta^2 \Omega}{4 \pi^2+T^2\Delta^2 }\right)}
}.
\label{eq:absprobp}
\end{equation}
This probability presents a maximum for $E=\Omega$,
\begin{equation}
P\bigg|_{E=\Omega} \simeq \frac{1   
}{
 \frac{\lambda^2 T \sqrt{2}}{16 \pi}   p(\Omega)  +\frac{4\pi^2 + T^2 \Delta^2}{\lambda^2 T^2 \Delta}
\frac{ 1}{\sqrt{2} p\left(  \Omega\right)}
}. \label{eq:absprobpmax}
\end{equation}

For an evanescent incoming particle,
the amplitudes of the processes (a), (b) and (c) differ from those of a propagating incoming particle by a phase. This is clear by comparison of expression \eqref{eq:propinint} with expression \eqref{eq:evaninint}. This means that the terms [a], [b] and [c] take the exact same form,
\begin{align}
[a]_{\text{evanescent}} &= [a]_{\text{propagating}}, \\
[b]_{\text{evanescent}} &= [b]_{\text{propagating}},\\
[c]_{\text{evanescent}} &= [c]_{\text{propagating}}.
\end{align}
The difference between the propagating and the evanescent sector occurs in term [m], as previously noted. In the evanescent case this is expression \eqref{eq:[m]} with $c=\im$. Appendix~\ref{sec:app-ap-m-ev} presents the calculation of [m], yielding
\begin{multline}
\rho_{\text{c}}[O_{g\to g}](\Psi^{\Lin}_{E,\Delta}\tens\Psi^{\Lout}_{E,\Delta}) + \text{c.c.} \\
  =  -  \im \frac{\lambda^2  \Delta^2 T^2}{ 4 \pi^2 + T^2 \Delta^2}  |c_{E,\Delta}|^2 e^{-\frac{2\pi T^2 }{4 \pi^2 + T^2 \Delta^2} (E-\Omega)^2}    
p\left(  \frac{4 \pi^2E + T^2 \Delta^2 \Omega}{4 \pi^2 + T^2 \Delta^2}\right)
 \mathrm{erf} \left( \im \frac{T}{ \sqrt{2 \pi}} \frac{4\pi^2(E-\Omega)}{4\pi^2+T^2 \Delta^2}  \right).
\end{multline}
Finally, the absorption probability takes the following form,
\begin{multline}
P \simeq \\
\frac{
1
}{
1 
+
\frac{\lambda^2 T \sqrt{2} \pi}{16 \pi^2 }  
p(\Omega)  
 \left[ 1 -
\mathrm{erf}^2 \left( \im \frac{T}{2 \sqrt{2 \pi}}\left[ \frac{4\pi^2(E-\Omega)}{4\pi^2+T^2 \Delta^2}  \right]\right)\right]
 -  \im   \mathrm{erf} \left( \im \frac{T}{ \sqrt{2 \pi}} \frac{4\pi^2(E-\Omega)}{4\pi^2+T^2 \Delta^2}  \right)
+
\frac{4\pi^2 + T^2 \Delta^2}{\sqrt{2} \lambda^2 T^2 \Delta}  
\frac{ e^{   \frac{ 2\pi T^2 (E-\Omega)^2}{4 \pi^2 +T^2 \Delta^2} }
}{ p\left(  \frac{4 \pi^2E + T^2 \Delta^2 \Omega}{4 \pi^2 + T^2 \Delta^2}\right)}
} .
\label{eq:absprobe}
\end{multline}
This probability presents a maximum for $E=\Omega$,
\begin{equation}
    P\bigg|_{E=\Omega} \simeq 
    \frac{1}{
    1 +\frac{\lambda^2 T \sqrt{2} \pi}{16 \pi^2 }  p(\Omega)  
    +\frac{4\pi^2 + T^2 \Delta^2}{\sqrt{2} \lambda^2 T^2 \Delta p\left(   \Omega\right)}
    } .
\label{eq:absprobemax}
\end{equation}

\subsection{Results}

\begin{figure}
    \begin{center}
    \includegraphics[width=0.7\textwidth]{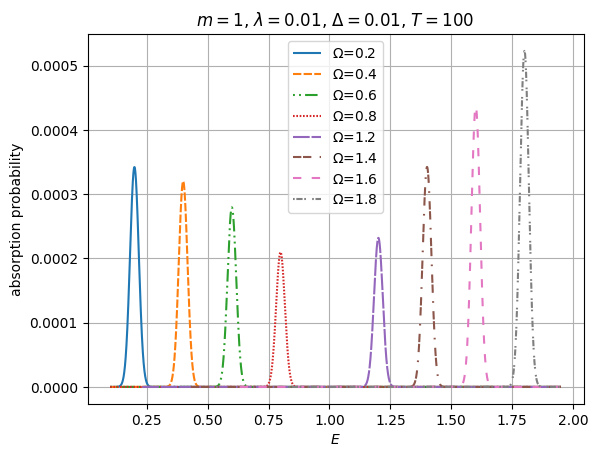}
    \caption{Absorption probability of an incoming particle as a function of the particle energy, for different detector gap energies at $\lambda=0.01$, $\Delta=0.01$, and $T=100$.}
    \label{fig:ap-ev}
    \end{center}
\end{figure}

We start by considering the absorption probability as a function of the energy of the incoming particle, see Figure~\ref{fig:ap-ev}. As expected, the absorption probability is peaked at the detector energy gap, with a Gaussian spread around it. In contrast to the case of spontaneous emission (e.g., Figure~\ref{fig:es}, right-hand side), this spread is not only caused by lack of adiabaticity (finite $T$), but also by the spread $\Delta$ explicitly introduced for the incoming wave packet. Another characteristic behavior that can be read off from Figure~\ref{fig:ap-ev} is the suppression of the absorption probability when the detector gap energy approaches the field mass, due to the square-root of momentum factor in the amplitude \eqref{eq:expra-p2}.

\begin{figure}
    \includegraphics[width=0.50\textwidth]{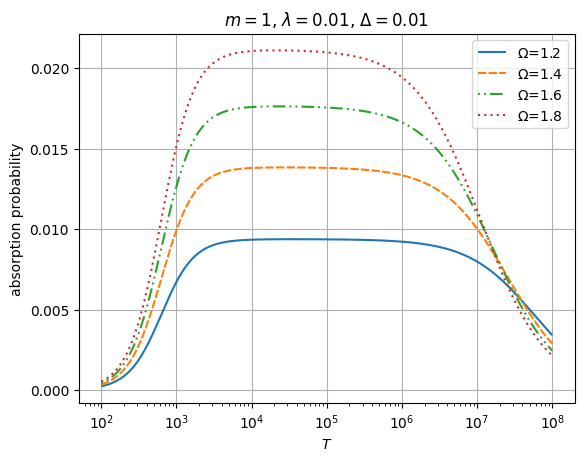}%
    \includegraphics[width=0.50\textwidth]{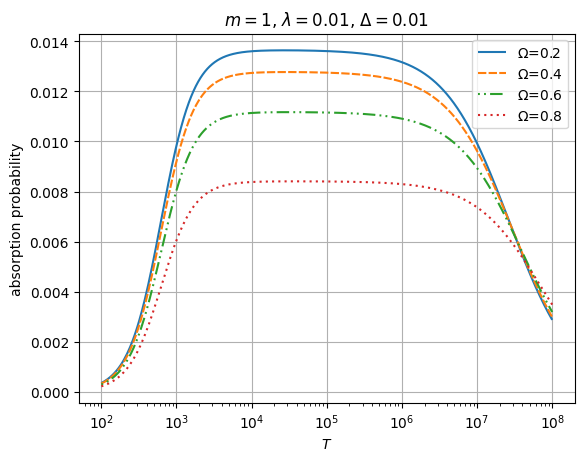}
    \caption{The absorption probability of an incoming particle with energy $E=\Omega$ is represented as a function of the characteristic time $T$ for the propagating sector (left-hand side) and the evanescent sector (right-hand side). Here, $\lambda=0.01$ and $\Delta=0.01$.}
    \label{fig:absorptionT}
\end{figure}

To explore the dependence of the absorption probability on the other variables we fix in the following the energy of the incoming particle to be equal to the detector gap energy. That is, we fix the particle energy so that the absorption probability is maximized. The dependence of this maximal absorption probability on the characteristic time $T$ is shown in Figure~\ref{fig:absorptionT}. The graph on the left-hand side shows the propagating sector, the one on the right-hand side the evanescent sector. Apart from the already noted decrease of the probability, in both sectors, when the detector energy gap approaches the field mass, we can read off the following. In the regime of small characteristic time $T$, the absorption probability increases markedly with $T$. Process (a), the possible absorption of the particle by the detector in the ground state dominates. At large $T$, on the other hand, we see a clear decay of the probability that the particle has been absorbed. Process (b) dominates here. That is, it is increasingly likely that the particle was absorbed by the detector and moreover has already been reemitted. For intermediate times, there appears to be a plateau, signaling an equilibrium of the likelihoods that the particle is just being absorbed or already being reemitted.

\begin{figure}
    \includegraphics[width=0.5\textwidth]{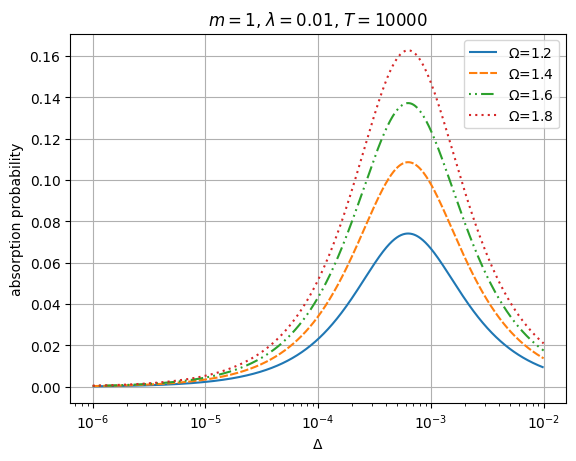}%
    \includegraphics[width=0.5\textwidth]{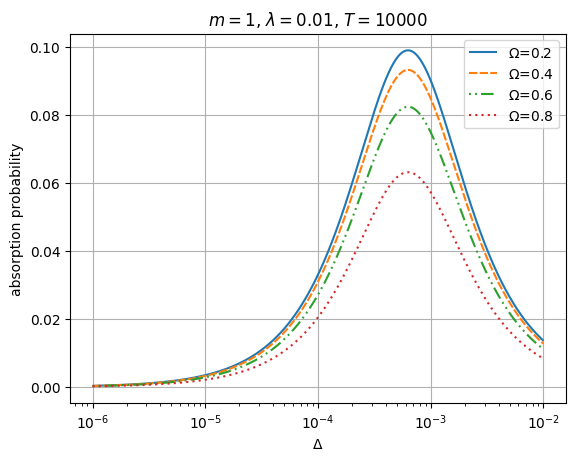}
    \caption{The absorption probability of an incoming particle with energy $E=\Omega$ is represented as a function of the energy spread $\Delta$ for the propagating sector (left-hand side) and the evanescent sector (right-hand side). Here, $\lambda=0.01$ and $T=10000$.}
    \label{fig:absorptionD}
\end{figure}

The maximal absorption probability as a function of the energy spread $\Delta$ of the incoming particle is shown in Figure~\ref{fig:absorptionD}. For relatively large values of $\Delta$ the absorption probability decreases. This is because more and more parts of the incoming wave packet correspond to energies increasingly different from the detector energy gap, moving them to the outer parts of the Gaussian peak of the absorption amplitude \eqref{eq:expra-p2}. From this point of view it might seem surprising that we also see a strong suppression of the absorption probability when $\Delta$ becomes very small, i.e., the particle energy becomes very peaked on the detector gap energy. Recall, however, that this is the regime where process (c) comes to dominate. That is, a particle with a more and more peaked energy is more and more likely to completely miss the detector, independent even of the detector gap energy. The physical explanation is that the particle's wave packet is increasingly delocalized in space, making it more difficult to deposit energy at the detector's location.


\section{Discussion and Outlook}
\label{sec:outlook}

In the present section we provide further comment on results, on methods used, and on directions for future research.
The first and foremost result of the present work is that evanescent particles behave in many ways just like ordinary propagating particles. In particular, they can be emitted and absorbed by a detector. Crucially, this process is accompanied by an exchange of energy. For propagating particles of a massive field, this energy is necessarily larger than the mass of the field. In contrast, for evanescent particles this energy is necessarily less than the mass of the field, but still positive and quantized. This result is embedded in a broader study of the interaction of a stationary UDW detector with the particle content of a surrounding massive quantum field. We study in some detail the emission spectrum (Section~\ref{sec:espectrum}), spontaneous emission probability (Section~\ref{sec:eprob}) and absorption probability (Section~\ref{sec:aprob}) of a UDW detector. In all cases evanescent particles play an important role and are part of our predictions.

The occurrence of evanescent particles has not been predicted in any previous work on the interaction of a UDW detector with a massive Klein-Gordon field. Before going into more details on our study, it is thus important to explain very clearly in which sense our predictions should or should not agree with the previous literature and why.
As recalled in the introduction, evanescent particles do not form part of the Hilbert space of a quantum field theory of massive particles on a spacelike hypersurface, but they do on a timelike hypersurface. This is a crucial ingredient when we discuss the description of the interaction of a UDW detector with the quantum field through two different pictures: the temporal vs.\ the radial picture. In the temporal picture we prepare an initial state of detector and field, let them interact for a certain time, and then measure their states. In the asymptotic case of infinite time this is the S-matrix description of the interaction. In contrast, in the radial picture preparation and measurement take place at a fixed distance from the detector, i.e., on a sphere of fixed radius with the detector at the center, at all times. As recalled in Section~\ref{sec:equivalence}, both pictures become strictly equivalent when this distance is taken to infinity \cite{CoOe:spsmatrix,CoOe:smatrixgbf}. However, when the distance is kept finite, additional degrees of freedom of the field become accessible, that are not present in the temporal picture. These manifest as evanescent particles.
Formalisms aside, the crucial point is that the radial picture (at finite radius) corresponds to a \emph{different experimental regime} from the temporal (S-matrix) picture. The key difference is the assumption in the temporal picture that there is nothing in the universe except for the detector and the field, during the whole time of interaction.\footnote{Strictly speaking, we know that some other sufficiently far away particles may not matter, due to cluster decomposition.} In the (finite) radial picture on the other hand there might well be other sources or sinks for the field present, at distance from the detector larger than the fixed radius. Crucially, our results do not depend directly on the presence or not of such sources or sinks. Instead, we condition on the presence or not of particles at the fixed finite radial distance from the detector. Now, say, the presence of an outgoing evanescent particle there might well be physically possible only if there is a device present further out, acting as a sink. Indeed, the absence of evanescent particles in the temporal picture tells us essentially that this must be the case. Broadly speaking, the radial picture is applicable when we allow to perform control or measurement tasks at finite distance from the detector during the experiment. If we imagine the UDW detector replaced by some generic interaction region, this scenario is in fact a more accurate modeling of how particle physics experiments are actually performed (think of the LHC) as compared to the temporal S-matrix picture. In any case, it is clear that we should expect our results (for the finite radial picture) to be different from those of the previous literature (for the temporal picture).

On the other hand, recall from Section~\ref{sec:equivalence} that when we exclude the evanescent sector in the radial picture (by fixing the state on the evanescent sector of the Hilbert space to be the vacuum), we recover equivalence to the temporal picture, even at finite radius. That is, of course, as long as we assume that there is no source or sink (or interaction term) outside of the sphere of fixed radius. This means that all our results of Sections~\ref{sec:espectrum}, \ref{sec:eprob} and \ref{sec:aprob} are valid for the temporal picture, as long as we cut out the parts pertaining to the evanescent sector. There are some interesting caveats to this statement that we discuss further along.

If the detector energy gap is larger than the mass of the field, the emission spectrum of a UDW detector is essentially a Gaussian, peaked at this gap energy (Section~\ref{sec:espectrum}). This is in accordance with expectations based on the previous literature, where mostly the case of a massless field was considered. The same behavior is observed, however, if the gap energy is below the field mass, see Figure~\ref{fig:es}. In this case the emitted particles are evanescent rather than propagating. Of course, as explained previously, this is only valid in the radial picture. We can also read off the corresponding result in the temporal or S-matrix picture. In that case the plots are identical\footnote{There is a small correction to normalization when the gap energy $\Omega$ is close to the mass $m$.} to the ones presented in the range $E>m$, while simply being cut off at $E<m$. (Here $m=1$.) It is interesting to note that the spectra (in the radial picture) look essentially symmetrical around $E=m$. That is, replacing $\Omega$ with $m-\Omega$ leads essentially to a mirror image spectrum. Moreover, there is a marked suppression of the spectra at $E=m$ arising from a square-root dependence of the amplitude on the momentum $p(E)=\sqrt{|E^2-m^2|}$ that vanishes just at $E=m$.

For the total probability of particle emission by an excited detector, it is clear that this increases with the time $T$ the detector is turned on (Section~\ref{sec:eprob}). The increase is at first approximately linear, if we exclude the regime of very low $T$ (because of the lack of adiabaticity of the detector switching in that regime.) Then a decay law qualitatively similar to, but different from, an exponential decay law takes over, with saturation at unit probability, see Figure~\ref{fig:sep-T}. For fixed time $T$, the probability increases with the energy of the detector gap, above the field mass. In the linear regime the probability is simply proportional to the particle momentum. This can be appreciated in Figure~\ref{fig:sep-Omega}. In particular, detector decay is almost completely prohibited at $\Omega=m$, with the probability dropping essentially to zero. This is the behavior both in the radial and in the temporal picture. The probability increases again when $\Omega$ drops below the field mass $m$, as the left-hand side ($\Omega<m$) of the plot in Figure~\ref{fig:sep-T} shows. This part is only valid in the radial picture. In the temporal picture, the probability is essentially zero at $\Omega<m$. However, this is not exactly so. What is more, for $\Omega$ near to $m$ the two pictures yield different results even for $\Omega>m$, see Figure~\ref{fig:sep-radtemp}. That is, there is a spillover effect of the evanescent sector into the propagating sector. With increasing adiabaticity of the switching (increasing $T$), this effect becomes confined to smaller and smaller neighborhoods of $\Omega=m$.

The probability of particle absorption by a UDW detector in the ground state is more complicated to determine as in addition to the pure absorption process, it is also possible that an absorbed particle is reemitted or that the particle "misses" the detector outright (Section~\ref{sec:aprob}). In terms of energy, the sensitivity of the detector is peaked around the detector gap energy, as expected, see Figure~\ref{fig:ap-ev}. What is new here is that this absorption is also possible at $\Omega<m$, with the particles absorbed in this case being evanescent ones, only present in the radial picture, not in the temporal one. It is then interesting to focus on the peak absorption probability, i.e., setting the particle energy equal to the energy gap. As expected from the previous results, this probability increases with $\Omega$ going away from $m$, in both directions, in the propagating ($\Omega>m$) and in the evanescent sector ($\Omega<m$). At $\Omega=m$ we again see a strong suppression, as in the case of emission. The behavior of the absorption probability on detector time is determined by two competing processes, see Figure~\ref{fig:absorptionT}. At short times the probability rises as the dominant process is the absorption of the particle while at long times the probability goes down again, eventually approaching zero, as reemission dominates. Another interesting trade-off can be seen in Figure~\ref{fig:absorptionD}. We find that a particle with a precise sharp energy (and thus sharp momentum) cannot be absorbed by the detector. This can be explained due to the complete delocalization stemming from the uncertainty relation. The probability for the particle to hit the point-localized detector drops to zero. To deal with this, we introduce wave functions with a Gaussian energy spread $\Delta$ around a central energy $E$. The absorption probability going to zero when the spread $\Delta$ goes to zero is the tail on the left-hand side in the plots of Figure~\ref{fig:absorptionD}. On the other hand, when the spread becomes large, more and more parts of the particle wave function correspond to energies increasingly detuned from the detector energy gap $\Omega$, where the detector sensitivity drops leading to a falloff in probability. That is seen on the right-hand side in the plots of Figure~\ref{fig:absorptionD}. These phenomena occur equally in the radial picture (both plots in Figures~\ref{fig:absorptionT} and \ref{fig:absorptionD}) as in the temporal picture (left-hand plots only in Figures~\ref{fig:absorptionT} and \ref{fig:absorptionD}).

We proceed to take a step back and look at the classical theory of evanescent modes and its \emph{quantization}.
Classically, the lack of oscillatory behavior means that we cannot associate a direction of propagation with evanescent modes by following the motion of maxima and minima of wave amplitudes in space as we do for propagating modes. This has led in \cite{CoOe:evanescent} and \cite{Oe:quanthcyl} to an ambiguity, not of the quantization scheme itself, but of the interpretation of one resulting binary degree of freedom. This issue was extensively discussed in both papers. In the present work we show that focusing on the flow of energy rather than the shape of the amplitude, does lead to a notion of spatial directionality also for evanescent waves. More precisely, measuring the flux of energy through the timelike hypercylinder via the energy-momentum tensor does lead to a separation of the space of evanescent modes into complementary subspaces of modes that carry energy from the exterior to the interior (incoming), or the other way round (outgoing). This separation thus precisely mirrors the separation existing in the propagating sector, see Section~\ref{sec:inout}. This resolves the ambiguity and establishes a correspondence to the propagating sector, at the classical level.

Working out the vertex amplitudes for the interaction of a UDW detector with a single particle of the Klein-Gordon field makes it clear that evanescent particles must behave in substantially the same way as propagating ones (Section~\ref{sec:emabs}). The formulas for the amplitudes of both types of particles can be brought into an identical form, up to a phase factor. What is more, this correspondence relies on the separation of modes into incoming vs.\ outgoing modes in both sectors alike. In other words, the \emph{quantized} incoming (outgoing) modes in the evanescent sector behave just like the \emph{quantized} incoming (outgoing) modes in the propagating sector, which are well understood. Since the quantization prescription for the evanescent sector (Section~\ref{sec:quantev}) is substantially different from the one for the propagating sector (Section~\ref{sec:quantprop}) this was far from obvious. In particular, incoming and outgoing modes in the evanescent sector conserve their classical characterization after quantization, just like modes in the propagating sector. We emphasize that this was not an input requirement when the \emph{twisted Kähler quantization} applied on the evanescent sector in Section~\ref{sec:quantev} was first designed in \cite{Oe:quanthcyl}. It is thus another significant result of the present work.

The methods used to describe the UDW detector, the massive Klein-Gordon field and their interaction are to a large extend based on standard tools from quantum mechanics and quantum field theory. This is particularly the case of our discussion of the temporal picture. To make sense of the radial picture on the other hand, the more powerful methods of general boundary quantum field theory are necessary \cite{Oe:gbqft,Oe:holomorphic,Oe:feynobs}, in particular for the novel twisted Kähler quantization \cite{CoOe:locgenvac} on timelike hypersurfaces (as applied in Section~\ref{sec:quanthypcyl}). The probability interpretation rests ultimately on the positive formalism \cite{Oe:posfound}, although a partial version adapted to scattering processes in the radial picture is already contained in \cite{Oe:kgtl}.

One aspect of our treatment of the UDW detector and its interaction with the quantum field that we would like to comment on further is our renormalization procedure (Section~\ref{sec:renorm}). As is well known, e.g.\ \cite{takagiVacuumNoiseStress1985,hummerRenormalizedUnruhDeWittParticle2016}, proceeding with a standard quantization of the detector-field interaction leads to divergent amplitudes at order higher than one of the perturbation theory. This is dealt with in the literature either by only ever considering perturbation theory at first order, or by smearing out the interaction in space, effectively giving the detector a nonzero size (as in Unruh's original paper \cite{Unr:bhevap}), which makes the higher order contributions finite. However, the latter does not guarantee convergence of the perturbation expansion. On the other hand, from a QFT perspective the divergences of the pointlike detector can be ascribed to a detector self-interaction mediated by the quantum field. This self-interaction may be described by replacing the original bare propagator for the detector with the \emph{complete} propagator, that includes all self-interaction diagrams (like that of Figure~\ref{fig:propvertexrenorm}.a). Similarly, the bare vertex (Figure~\ref{fig:vertices}) is replaced by the \emph{complete} vertex, that includes all self-interaction diagrams (like that of Figure~\ref{fig:propvertexrenorm}.b). Of course, with the bare propagator and vertex as given, the complete propagator and vertex would be infinite, precisely due to the divergences. Our \emph{renormalization procedure} now consists in declaring that the original propagator and vertex are to be considered not the bare ones, but already the complete ones. This is consistent, if we exclude any further self-interaction of these complete objects. Technically this is implemented precisely by \emph{normal ordering} of the detector observable. An advantage of this procedure is that for a state with particle number $n$, only terms of order up to $n$ in the perturbation expansion contribute, which allows us to present "exact" results. Note that our procedure is different from the more conventional normal ordering at the level of the Hamiltonian, as applied for example for certain detector models with higher order coupling to the field \cite{hummerRenormalizedUnruhDeWittParticle2016}.

In the present work we have considered a single detector and shown that it can absorb and emit evanescent particles, very much in the same way as propagating particles. A next step would be to study multiple detectors. Using general boundary quantum field theory we can insert a timelike hypersurface between detectors to "intercept" both propagating and evanescent particles between the detectors. This should shed further light on their similarities and differences. In particular, we expect to see an exponential decline in the interaction with increasing distance in the evanescent case.

The scalar Klein-Gordon field used in the present work is a convenient theoretical vehicle due to its simplicity. However, if we want to get closer to experimental predictions and statements about the real world, we should study the electromagnetic field. Crucially, although massless, this also admits evanescent modes, and thus provides an important motivation for the study of evanescent particles. Versions of the UDW detector interacting with an electromagnetic field have been known for quite some time \cite{takagiVacuumNoiseStress1985}. Ultimately this goes back to descriptions of the interaction of the electromagnetic field with matter in quantum optics \cite{glauberCoherentIncoherentStates1963}. A next step is thus to repeat a study like the present one, but with the massive Klein-Gordon field replaced by the massless electromagnetic field.

In the present paper we have only considered a detector at rest. However, essentially all of the machinery we have developed to describe the UDW detector and its interaction with the quantum field (Sections~\ref{sec:udwhypcyl}--\ref{sec:feynrenorm}) is applicable to a detector with a quite arbitrary trajectory. We thus hope our work to prove useful for studies with a moving detector also.

\subsection*{Acknowledgments}

This work was partially supported by UNAM-PAPIIT project grant IN106422.
The work of A.~Z.\ was partially supported by the National Science Centre in Poland under the research grant Maestro (2021/42/A/ST2/0035).

\renewcommand{\thesubsection}{\arabic{subsection}.}
\numberwithin{equation}{section}

\begin{appendices}


\section{Regularity of the switching function}
\label{sec:regularity}

An obvious first requirement for the well-definedness of probabilities is the finiteness of the vertex amplitudes \eqref{eq:emampl} and \eqref{eq:absampl}. This is equivalent to the integrability of the switching function $\chi$. For later use we point out that integrability implies,
\begin{equation}
    \lim_{E\to\infty}\int_{-\infty}^{\infty}\xd \tau\,
    \chi(\tau) e^{\im E\tau} =0 .
    \label{eq:conv1}
\end{equation}
This is an elementary property of the Fourier transform. Alternatively, this is easy to derive by approximating the switching function with characteristic functions of finite intervals.

A further requirement in order to obtain well-defined probabilities or probability densities for emission and absorption processes is normalizability. In the present context this means that the energy integrals in the denominators of expressions \eqref{eq:specexpem}, \eqref{eq:probexpse} and \eqref{eq:absprob} have to be finite. It is sufficient to consider the case of emission. That is, we consider the integral over all energies of the modulus square of the amplitude \eqref{eq:amplemitpint} for the emission of a particle of energy $E$. For convergence only the high-energy contribution of the integrand is relevant, so we can set $p(E)\approx E$, and $E-\Omega\approx E$, and reduce this to the question of convergence of the following integral:
\begin{equation}
    \int_{m}^{\infty}\xd E\, E \left|\int_{-\infty}^{\infty}\xd \tau\,
    \chi(\tau) e^{\im E\tau}\right|^2 .
    \label{eq:probint}
\end{equation}
Again, it is convenient to characterize the convergence of this integral in terms of properties of the Fourier transform of $\chi$, as there is a relation between the decay properties of the latter and differentiability of $\chi$. It is easy to see that a \emph{necessary condition} for convergence is the following:
\begin{equation}
    \lim_{E\to\infty} E \int_{-\infty}^{\infty}\xd \tau\,
    \chi(\tau) e^{\im E\tau} =0 .
    \label{eq:conv2}
\end{equation}
This is satisfied if $\chi$ is in addition differentiable and $\chi'$ is integrable, since,
\begin{equation}
    E \int_{-\infty}^{\infty}\xd \tau\, \chi(\tau) e^{\im E\tau}
    =-\im\int_{-\infty}^{\infty}\xd \tau\, \chi(\tau) \frac{\xd}{\xd\tau} e^{\im E\tau}
    =-\im \chi(\tau)e^{\im E\tau}\bigg|_{-\infty}^{\infty}
     +\im \int_{-\infty}^{\infty}\xd \tau\, \chi'(\tau) e^{\im E\tau} .
    \label{eq:convderiv}
\end{equation}
Then, \eqref{eq:conv2} follows with \eqref{eq:conv1}. On the other hand, a \emph{sufficient condition} for the convergence of \eqref{eq:probint} is
\begin{equation}
    \lim_{E\to\infty} E^2 \int_{-\infty}^{\infty}\xd \tau\,
    \chi(\tau) e^{\im E\tau} =0 .
    \label{eq:conv3}
\end{equation}
Multiplying \eqref{eq:convderiv} with $E$ we find that we require
\begin{equation}
    \lim_{E\to\infty} E \int_{-\infty}^{\infty}\xd \tau\,
    \chi'(\tau) e^{\im E\tau} =0 .
    \label{eq:conv3b}
\end{equation}
Repeating the previous argument for $\chi'$ in place of $\chi$ we see that $\chi$ needs to be twice differentiable and $\chi''$ needs to be integrable.


\section{Quantization: additional relations}
\label{sec:extraquant}

The present appendix provides additional expressions arising in the quantization considered in Section~\ref{sec:quanthypcyl} that are required for some calculations.
The real structure $\alpha$ is given as follows:
\begin{gather}
    (\alpha(\xi))_{E,\ls,\ms}^{\Lout}=(-1)^{\ls}\im\, \overline{\xi_{E,\ls,\ms}^{\Lcin}},\;
    (\alpha(\xi))_{E,\ls,\ms}^{\Lcout}=(-1)^{\ls}\im\, \overline{\xi_{E,\ls,\ms}^{\Lin}}, \nonumber\\
    (\alpha(\xi))_{E,\ls,\ms}^{\Lin}=(-1)^{\ls}\im\, \overline{\xi_{E,\ls,\ms}^{\Lcout}},\;
    (\alpha(\xi))_{E,\ls,\ms}^{\Lcin}=(-1)^{\ls}\im\, \overline{\xi_{E,\ls,\ms}^{\Lout}} .
    \label{eq:alpha}
\end{gather}
The linear mapping $I^{\Le}:L^{\Le}\to L^{\Le,\alpha}$ and its complexification is given as follows:
\begin{gather}
  \left(I^{\Le}(\phi)\right)^{\Lout}_{E,\ls,\ms}
  =\frac{1}{\sqrt{2}}\left(\phi^{\Lout}_{E,\ls,\ms}
  +(-1)^{\ls}\im \phi^{\Lin}_{E,\ls,\ms}\right), \quad
  \left(I^{\Le}(\phi)\right)^{\Lcout}_{E,\ls,\ms}
  =\frac{1}{\sqrt{2}}\left(\phi^{\Lcout}_{E,\ls,\ms}
  +(-1)^{\ls}\im \phi^{\Lcin}_{E,\ls,\ms}\right), \nonumber \\
  \left(I^{\Le}(\phi)\right)^{\Lin}_{E,\ls,\ms}
  =\frac{1}{\sqrt{2}}\left(\phi^{\Lin}_{E,\ls,\ms}
  +(-1)^{\ls}\im \phi^{\Lout}_{E,\ls,\ms}\right), \quad
  \left(I^{\Le}(\phi)\right)^{\Lcin}_{E,\ls,\ms}
  =\frac{1}{\sqrt{2}}\left(\phi^{\Lcin}_{E,\ls,\ms}
  +(-1)^{\ls}\im \phi^{\Lcout}_{E,\ls,\ms}\right) .
  \label{eq:idralphaexp}
\end{gather}
The polarizations in the propagating and evanescent sectors induce complex structures on $L^{\Lp}$ and $L^{\Le,\alpha}$ respectively,
\begin{gather}
  (J^{\Lp}\phi)^{\Lout}_{E,\ls,\ms}=\im \phi^{\Lout}_{E,\ls,\ms},\quad
  (J^{\Lp}\phi)^{\Lcout}_{E,\ls,\ms}=-\im\phi^{\Lcout}_{E,\ls,\ms},\nonumber \\
  (J^{\Lp}\phi)^{\Lin}_{E,\ls,\ms}=-\im\phi^{\Lin}_{E,\ls,\ms},\quad
  (J^{\Lp}\phi)^{\Lcin}_{E,\ls,\ms}=\im \phi^{\Lcin}_{E,\ls,\ms}, \\
%
  (J^{\Le}\phi)^{\Lout}_{E,\ls,\ms}=(-1)^{\ls}\phi^{\Lin}_{E,\ls,\ms},\quad
  (J^{\Le}\phi)^{\Lcout}_{E,\ls,\ms}=-(-1)^{\ls}\phi^{\Lcin}_{E,\ls,\ms},\nonumber \\
  (J^{\Le}\phi)^{\Lin}_{E,\ls,\ms}=-(-1)^{\ls}\phi^{\Lout}_{E,\ls,\ms},\quad
  (J^{\Le}\phi)^{\Lcin}_{E,\ls,\ms}=(-1)^{\ls}\phi^{\Lcout}_{E,\ls,\ms} .
\end{gather}
The composition of the map $I^{\Le}$ with the inner product \eqref{eq:ebilin} in the evanescent sector recovers an expression reminiscent of the inner product of the propagating sector \eqref{eq:pbilin},
\begin{equation}
  \{I^{\Le}(\phi),I^{\Le}(\xi)\}^{\Le}
  =\int_{0}^{m}\xd E\frac{p}{2\pi}\sum_{\ls,\ms}
  \left(\xi_{E,\ls,\ms}^\Lout\phi_{E,\ls,\ms}^\Lcout
  + \xi_{E,\ls,\ms}^\Lcin\phi_{E,\ls,\ms}^\Lin\right) .
  \label{eq:Ibilin}
\end{equation}
For both, propagating and evanescent sector, we have
\begin{equation}
  J L_M^{\C}=\{\phi\in L^{\C}: \phi^{\Lin}_{E,\ls,\ms}=-\phi^{\Lout}_{E,\ls,\ms}, \phi^{\Lcin}_{E,\ls,\ms}=-\phi^{\Lcout}_{E,\ls,\ms}\} .
\end{equation}
This gives rise to a direct sum decomposition $L^{\C}=L_M^{\C}\oplus J L_M^{\C}$. The map $u:L\to L$ is defined as the identity on $L_M^{\C}$ and minus the identity on $J L_M^{\C}$ \cite{Oe:quanthcyl}. Here, it takes the form,
\begin{gather}
  (u(\phi))^{\Lout}_{E,\ls,\ms}=\phi^{\Lin}_{E,\ls,\ms},\quad
  (u(\phi))^{\Lcout}_{E,\ls,\ms}=\phi^{\Lcin}_{E,\ls,\ms},\nonumber \\
  (u(\phi))^{\Lin}_{E,\ls,\ms}=\phi^{\Lout}_{E,\ls,\ms},\quad
  (u(\phi))^{\Lcin}_{E,\ls,\ms}=\phi^{\Lcout}_{E,\ls,\ms} .
\end{gather}
Key properties of this map are
\begin{equation}
  \{u(\xi),u(\eta)\}=\{\eta,\xi\}, \qquad
  \overline{\{\xi,\eta\}}=\{\alpha(\eta),\alpha(\xi)\} .
\end{equation}
For $\xi\in L^{\Lp}\oplus L^{\Le,\alpha}$ we have,
\begin{equation}
  \alpha(\xi^{\Lint})=(u(\xi))^{\Lint} .
\end{equation}
Remarkably, the map $\alpha$ can be expressed in terms of complex conjugation and $u$, with a parity dependence on angular momentum,
\begin{equation}
   (\alpha(\xi))^{\bullet}_{E,\ls,\ms}=(-1)^{\ls}(\overline{u(\xi)})^{\bullet}_{E,\ls,\ms} .
\end{equation}


\section{Absorption probability}
\label{sec:app2}
In this appendix we derive the expression of the amplitude of the processes (a) and (b) that form part of the calculation of the absorption probability of an incoming propagating particle.

\subsection{Process (a)}
\label{sec:app-ap-a}
The substitution of expression \eqref{eq:expra-p2} into \eqref{eq:expra-p} leads to
\begin{multline}
   \rho[O_{g\to e}](\Psi^{\Lin}_{E,\Delta}) \overline{\rho[O_{g\to e}](\Psi^{\Lin}_{E,\Delta})} \\
    = |c_{E,\Delta}|^2 \lambda^2 T^2 \int_m^\infty \xd E' \, \xd E''  \sqrt{p(E')p(E'')} e^{-\pi \frac{(E-E')^2+(E-E'')^2}{\Delta^2}} e^{- \frac{(E'- \Omega)^2 +(E''- \Omega)^2}{4 \pi} T^2 } .
 \end{multline}
By noticing that
\begin{equation}
e^{ -\pi  \left(\frac{E-E'}{\Delta} \right)^2 }  e^{-\frac{(E'-\Omega)^2}{4 \pi} T^2}
= e^{-\frac{\pi T^2 }{4 \pi^2 + T^2 \Delta^2} (E-\Omega)^2} 
e^{-\frac{4 \pi^2 + T^2 \Delta^2}{4\pi \Delta^2} \left( E' - \frac{4 \pi^2E + T^2 \Delta^2 \Omega}{4 \pi^2 + T^2 \Delta^2}\right)^2 }
\label{eq:id-exp}
\end{equation}
the product of amplitudes can be rewritten in the following form:
\begin{multline}
\rho[O_{g\to e}](\Psi^{\Lin}_{E,\Delta}) \overline{\rho[O_{g\to e}](\Psi^{\Lin}_{E,\Delta})} \\
=|c_{E,\Delta}|^2 \lambda^2 T^2    \exp \left(  - \frac{ 2\pi T^2 }{4 \pi^2 +T^2 \Delta^2}(E-\Omega)^2 \right)
\int_m^\infty \xd E' \, \xd E''  \sqrt{p(E')p(E'')} \\
 \exp \left( - \frac{4 \pi^2+T^2\Delta^2}{4 \pi \Delta^2} \left[\left(E'  - \frac{4 \pi^2 E + T^2 \Delta^2 \Omega}{4 \pi^2+T^2\Delta^2 }\right)^2 +\left(E''  - \frac{4 \pi^2 E + T^2 \Delta^2 \Omega}{4 \pi^2+T^2\Delta^2 }\right)^2 \right] \right) .
\end{multline}
Both integrals can be evaluated by using the Laplace method assuming that $\frac{4 \pi^2+T^2\Delta^2}{4 \pi \Delta^2} \gg1$,
\begin{multline}
\rho[O_{g\to e}](\Psi^{\Lin}_{E,\Delta}) \overline{\rho[O_{g\to e}](\Psi^{\Lin}_{E,\Delta})} \\
\simeq   \frac{\lambda^2 T^2 \Delta^2}{4\pi^2 + T^2 \Delta^2}  |c_{E,\Delta}|^2   \exp \left(  - \frac{ 2\pi T^2 }{4 \pi^2 +T^2 \Delta^2} (E-\Omega)^2\right) p\left( \frac{4 \pi^2 E + T^2 \Delta^2 \Omega}{4 \pi^2+T^2\Delta^2 }\right) .
\end{multline}

\subsection{Process (b)}
\label{sec:app-ap-b}
The amplitude of the process (b) is given by
\begin{align}
&\rho_{\text{c}}[O_{g\to g}](\Psi^{\Lin}_{E,\Delta}\tens\Psi^{\Lout}_{E'})\nonumber\\
 &=- \lambda^2 2 c_{E,\Delta} \int_m^{\infty} \xd E' e^{ -\pi  \left(\frac{E-E'}{\Delta} \right)^2} \nonumber\\
 & \quad
\int^{\infty}_{-\infty} \xd \tau  \, \chi(\tau) (\Phi^{\Lout,E'',0,0})^{\Lint}(\tau,\vec{0}) e^{-\im \Omega \tau} 
  \int^{\tau}_{-\infty} \xd {\tau'} \, \chi(\tau') (\Phi^{\Lin,E'',0,0})^{\Lint}(\tau',\vec{0}) e^{\im \Omega \tau'},\\
 &=-  \frac{\lambda^2 T}{8  \pi^2}  c_{E,\Delta} \int_m^{\infty} \xd E' 
e^{ -\pi  \left(\frac{E-E'}{\Delta} \right)^2}  \sqrt{p(E'')p(E')}e^{-\frac{(E'-\Omega)^2}{4 \pi^2} T^2} \nonumber\\
& \quad \left( 
\int^{\infty}_{-\infty} \xd \tau  \, \chi(\tau) e^{-\im (\Omega-E'') \tau} 
\mathrm{erf} \left( \frac{2 \pi \tau + \im T^2(E'-\Omega)}{2 \sqrt{\pi} T} \right)+ Te^{-\frac{(E''-\Omega)^2}{4 \pi} T^2}\right).
\label{eq:app-amp-b}
 \end{align}
 The integral over $\tau$ can be evaluated by expression 2.7.1.6 of \cite{Korotkov2020}, namely
 \begin{equation}
 \int_{-\infty}^{\infty} e^{-az^2+\beta z} \mathrm{erf} \left(a_1z+\beta_1 \right) = \sqrt{\frac{\pi}{a}} e^{\beta^2/4a} \mathrm{erf} \left( \frac{2a \beta_1 +a_1 \beta}{2 \sqrt{a^2+a a_1^2}} \right),
 \end{equation}
 with the following identifications:
 \begin{align}
 a= \frac{\pi}{T^2}, \quad \beta = -\im (\Omega-E''), \quad a_1 = \frac{\sqrt{\pi}}{T}, \quad \beta_1 = \im \frac{(E'-\Omega)T}{2 \sqrt{\pi}} .
 \end{align}
 Thus,
 \begin{equation}
 \int^{\infty}_{-\infty} \xd \tau  \, \chi(\tau) e^{-\im (\Omega-E'') \tau} \mathrm{erf} \left( \frac{2 \pi \tau + \im T^2(E'-\Omega)}{2 \sqrt{\pi} T} \right) 
 = T e^{ - \frac{(E''-\Omega)^2T^2}{4 \pi}} \mathrm{erf} \left( \im \frac{T}{2 \sqrt{2 \pi}} \left[E'+E''- 2 \Omega \right] \right) .
 \label{eq:int-Kor}
 \end{equation}
 Expression \eqref{eq:app-amp-b} becomes
 \begin{multline}
\rho_{\text{c}}[O_{g\to g}](\Psi^{\Lin}_{E,\Delta}\tens\Psi^{\Lout}_{E'})\\
 =-  \frac{\lambda^2 T^2}{8  \pi^2}  c_{E,\Delta}   e^{-\frac{(E''-\Omega)^2}{4 \pi} T^2} \int_m^{\infty} \xd E' 
e^{ -\pi  \left(\frac{E-E'}{\Delta} \right)^2}  \sqrt{p(E'')p(E')}e^{-\frac{(E'-\Omega)^2}{4 \pi^2} T^2} \\
\left( 
 \mathrm{erf} \left( \im \frac{T}{2 \sqrt{2 \pi}} \left[E'+E''- 2 \Omega \right] \right)
+1\right).
 \end{multline}
 The product of the exponentials appearing under the integration can we rewritten as in \eqref{eq:id-exp}. Then
 the Laplace method provides an approximation of the integral over $E'$, valid for $T\gg1$,
 \begin{multline}
\rho_{\text{c}}[O_{g\to g}](\Psi^{\Lin}_{E,\Delta}\tens\Psi^{\Lout}_{E'})\\
  \simeq -  \frac{\lambda^2  \Delta T^2}{4  \pi \sqrt{4 \pi^2 + T^2 \Delta^2}}  c_{E,\Delta}e^{-\frac{\pi T^2 }{4 \pi^2 + T^2 \Delta^2} (E-\Omega)^2}   
 \sqrt{p(E'')p\left(  \frac{4 \pi^2E + T^2 \Delta^2 \Omega}{4 \pi^2 + T^2 \Delta^2}\right)} e^{-\frac{(E''-\Omega)^2}{4 \pi} T^2}
\\
  \left( \mathrm{erf} \left( \im \frac{T}{2 \sqrt{2 \pi}}\left[ \frac{4\pi^2(E-\Omega)}{4\pi^2+T^2 \Delta^2}  + E''- \Omega\right]\right)+1\right).
 \end{multline}
 
 \subsection{Mixed term in the evanescent sector}
 \label{sec:app-ap-m-ev}

 The mixed term [m] in the evanescent sector is obtained from expression \eqref{eq:mterm},
 \begin{multline}
\rho_{\text{c}}[O_{g\to g}](\Psi_{\Phi^{\Lin}_{E,\Delta}}\tens\Psi_{u(\Phi^{\Lin}_{E,\Delta})})  + \text{c.c.} \\
= |c_{E,\Delta}|^2\int_0^m \xd E' \, \xd E'' \, e^{-\pi\left(\frac{E-E'}{\Delta}\right)^2} e^{-\pi\left(\frac{E-E''}{\Delta}\right)^2}
\left[ \rho_{\text{c}}[O_{g\to g}](\Psi^{\Lin}_{E'}\tens\Psi^{\Lout}_{E''})  + \text{c.c.} \right],
\end{multline}
where the amplitude appearing in the last parenthesis is
 \begin{align}
 & \rho_{\text{c}}[O_{g\to g}](\Psi^{\Lin}_{E'}\tens\Psi^{\Lout}_{E''})  \nonumber\\
 &=-2 \lambda^2\int^{\infty}_{-\infty} \xd \tau  \, \chi(\tau) e^{-\im \Omega \tau}  (\Phi^{\Lout,E',0,0})^{\Lint}(\tau,\vec{0}) 
\int^{\tau}_{-\infty} \xd {\tau'} \, \chi(\tau') e^{\im \Omega \tau'}  (\Phi^{\Lin,E'',0,0})^{\Lint}(\tau',\vec{0}), \\
 &=-\frac{2 \im}{(4 \pi)^2} \lambda^2 T \sqrt{p(E'')p(E')}e^{- \frac{(E'-\Omega)^2}{4 \pi} T^2} \nonumber\\
& 
 \left( \int^{\infty}_{-\infty} \xd \tau  \, \chi(\tau) e^{-\im( \Omega - E'') \tau}
 \mathrm{erf} \left( \frac{2 \pi \tau + \im T^2(E'-\Omega)}{2 \sqrt{\pi} T} \right) +T e^{- \frac{(E''-\Omega)^2}{4 \pi} T^2} \right) .
 \end{align}
 The integral over $\tau$ coincides with \eqref{eq:int-Kor}, leading to
 \begin{multline}
  \rho_{\text{c}}[O_{g\to g}](\Psi^{\Lin}_{E'}\tens\Psi^{\Lout}_{E''})  \\
  =-\frac{2 \im}{(4 \pi)^2} \lambda^2 T^2 \sqrt{p(E'')p(E')}e^{- \frac{(E'-\Omega)^2}{4 \pi} T^2}  e^{ - \frac{(E''-\Omega)^2T^2}{4 \pi}} 
 \left( \mathrm{erf} \left( \im \frac{T}{2 \sqrt{2 \pi}} \left[E'+E''- 2 \Omega \right] \right) +1 \right).
 \end{multline}
 Since $\im \, \mathrm{erf}(\im x)$ is real, the sum of the above amplitude with its complex conjugate results to be
 \begin{multline}
 \rho_{\text{c}}[O_{g\to g}](\Psi^{\Lin}_{E'}\tens\Psi^{\Lout}_{E''})  + \text{c.c.}\\
   =-\frac{ \im}{4 \pi^2} \lambda^2 T^2 \sqrt{p(E'')p(E')}e^{- \frac{(E'-\Omega)^2}{4 \pi} T^2}  e^{ - \frac{(E''-\Omega)^2T^2}{4 \pi}} 
 \mathrm{erf} \left( \im \frac{T}{2 \sqrt{2 \pi}} \left[E'+E''- 2 \Omega \right] \right) .
 \end{multline} 
To evaluate the integrals over $E'$ and $E''$ we express the integrand in terms of equality \eqref{eq:id-exp} and then apply the Laplace method, obtaining
  \begin{align}
&\rho_{\text{c}}[O_{g\to g}](\Psi_{\Phi^{\Lin}_{E,\Delta}}\tens\Psi_{u(\Phi^{\Lin}_{E,\Delta})})  + \text{c.c.} \nonumber\\
 &=-\im |c_{E,\Delta}|^2    \frac{ \lambda^2 T^2 \Delta^2}{4 \pi^2 + T^2 \Delta^2}  
  e^{-\frac{2\pi T^2 }{4 \pi^2 + T^2 \Delta^2} (E-\Omega)^2}
 p \left( \frac{4 \pi^2E + T^2 \Delta^2 \Omega}{4 \pi^2 + T^2 \Delta^2}\right)
 \mathrm{erf} \left( \im \frac{T}{ \sqrt{2 \pi}}  \frac{4 \pi^2(E-\Omega) }{4 \pi^2 + T^2 \Delta^2}  \right).
\end{align}

\end{appendices}

\newcommand{\eprint}[1]{\href{https://arxiv.org/abs/#1}{#1}}
\bibliographystyle{stdnodoi} 
\bibliography{stdrefsb,detector}
\end{document}